\documentclass[apj,iop,twocolappendix,numberedappendix]{emulateapj}

\usepackage{relsize}
\usepackage{color}
\usepackage{xcolor}
\usepackage{amsmath}
\usepackage{mathtools}
\usepackage{multirow}
\usepackage{hyperref}
\usepackage{graphicx}

\hypersetup{
     colorlinks   = true,
     citecolor    = blue,
     linkcolor    = blue,
     urlcolor    = blue
}

\newcommand{\sigrho}{\sigma_{\rho/\rho_0}}

\newcommand{\rflat}{R_\mathrm{flat}}

\newcommand{\mach}{\mathcal{M}}
\newcommand{\macha}{\mathcal{M}_\mathrm{A}}

\newcommand{\sfrff}{\eps_\mathrm{ff}}

\newcommand{\msol}{\mbox{$M_{\sun}$}}
\newcommand{\eps}{\epsilon}

\newcommand{\cs}{c_\mathrm{s}}

\newcommand{\va}{v_\mathrm{A}}
\newcommand{\ls}{\lambda_\mathrm{sonic}}

\newcommand{\km}{\mathrm{km}}
\newcommand{\pc}{\mathrm{pc}}

\newcommand{\s}{\mathrm{s}}
\newcommand{\yr}{\mathrm{yr}}

\newcommand{\Gauss}{\mathrm{G}}
\newcommand{\cm}{\mbox{cm}}
\newcommand{\g}{\mbox{g}}
\newcommand{\G}{\mbox{G}}

\newcommand{\brick}{G0.253+0.016}

\slugcomment{accepted for publication in ApJ, September 19, 2016}

\shorttitle{The CMZ cloud \brick}
\shortauthors{Federrath et al.}

\begin{document}

\title{The link between turbulence, magnetic fields, filaments, and star formation \\in the Central Molecular Zone cloud \brick}

\author{C.~Federrath\altaffilmark{1}, J.~M.~Rathborne\altaffilmark{2}, S.~N.~Longmore\altaffilmark{3}, J.~M.~D.~Kruijssen\altaffilmark{4,5}, J.~Bally\altaffilmark{6}, Y.~Contreras\altaffilmark{7}, R.~M.~Crocker\altaffilmark{1}, G.~Garay\altaffilmark{8}, J.~M.~Jackson\altaffilmark{9}, L.~Testi\altaffilmark{10,11,12}, A.~J.~Walsh\altaffilmark{13}}
\email{christoph.federrath@anu.edu.au}

\altaffiltext{1}{Research School of Astronomy and Astrophysics, Australian National University, Canberra, ACT~2611, Australia}
\altaffiltext{2}{CSIRO Astronomy and Space Science, P.O.~Box~76, Epping NSW, 1710, Australia}
\altaffiltext{3}{Astrophysics Research Institute, Liverpool John Moores University, IC2, Liverpool Science Park, 146~Brownlow Hill, Liverpool~L3~5RF, United Kingdom}
\altaffiltext{4}{Astronomisches Rechen-Institut, Zentrum f\"ur Astronomie der Universit\"at Heidelberg, M\"onchhofstra{\ss}e 12-14, 69120~Heidelberg, Germany}
\altaffiltext{5}{Max-Planck Institut f\"{u}r Astronomie, K\"{o}nigstuhl~17, 69117~Heidelberg, Germany}
\altaffiltext{6}{CASA, University of Colorado, 389-UCB, Boulder, CO~80309, USA}
\altaffiltext{7}{Leiden Observatory, Leiden University, PO~Box~9513, NL-2300 RA Leiden, the Netherlands}
\altaffiltext{8}{Departamento de Astronom\'ia, Universidad de Chile, Casilla 36-D, Santiago, Chile}
\altaffiltext{9}{Institute for Astrophysical Research, Boston University, Boston, MA~02215, USA}
\altaffiltext{10}{European Southern Observatory, Karl-Schwarzschild-Stra{\ss}e~2, D-85748 Garching bei M\"unchen, Germany}
\altaffiltext{11}{INAF-Arcetri, Largo E.~Fermi~5, I-50125 Firenze, Italy}
\altaffiltext{12}{Excellence Cluster Universe, Boltzmannstra{\ss}e~2, D-85748, Garching, Germany}
\altaffiltext{13}{International Centre for Radio Astronomy Research, Curtin University, GPO Box U1987, Perth WA~6845, Australia}

\begin{abstract}
Star formation is primarily controlled by the interplay between gravity, turbulence, and magnetic fields. However, the turbulence and magnetic fields in molecular clouds near the Galactic Center may differ substantially from spiral-arm clouds. Here we determine the physical parameters of the central molecular zone (CMZ) cloud \brick, its turbulence, magnetic field and filamentary structure. Using column-density maps based on dust-continuum emission observations with ALMA+\emph{Herschel}, we identify filaments and show that at least one dense core is located along them. We measure the filament width $W_\mathrm{fil}=0.17\pm0.08\,\pc$ and the sonic scale $\lambda_\mathrm{sonic}=0.15\pm0.11\,\pc$ of the turbulence, and find $W_\mathrm{fil}\approx\lambda_\mathrm{sonic}$. A strong velocity gradient is seen in the HNCO intensity-weighted velocity maps obtained with ALMA+Mopra. The gradient is likely caused by large-scale shearing of {\brick}, producing a wide double-peaked velocity PDF. After subtracting the gradient to isolate the turbulent motions, we find a nearly Gaussian velocity PDF typical for turbulence. We measure the total and turbulent velocity dispersion, $8.8\pm0.2\,\km\,\s^{-1}$ and $3.9\pm0.1\,\km\,\s^{-1}$, respectively. Using magnetohydrodynamical turbulence simulations, we find that {\brick}'s turbulent magnetic field $B_\mathrm{turb}=130\pm50\,\mu\Gauss$ is only $\lesssim1/10$ of the ordered field component. Combining these measurements, we reconstruct the dominant turbulence driving mode in {\brick} and find a driving parameter $b=0.22\pm0.12$, indicating solenoidal (divergence-free) driving. We compare this to spiral-arm clouds, which typically have a significant compressive (curl-free) driving component ($b>0.4$). Motivated by previous reports of strong shearing motions in the CMZ, we speculate that shear causes the solenoidal driving in {\brick} and show that this reduces the star formation rate (SFR) by a factor of $6.9$ compared to typical nearby clouds.
\end{abstract}

\keywords{\\Galaxy: center --- galaxies: ISM --- ISM: clouds --- magnetic fields --- stars: formation --- turbulence}

\section{Introduction}

Star formation powers the evolution of galaxies. However, the processes that control the conversion of gas into stars remain poorly understood. We now know that turbulence, magnetic fields and feedback are essential for regulating star formation in the Galactic disk, because gravity alone would produce stars at a $\sim\!100$ times higher rate than observed \citep{McKeeOstriker2007,PadoanEtAl2014,Federrath2015}. However, it is not so clear whether the same principles hold in the Central Molecular Zone---a much more extreme environment. For instance, despite the high gas densities and the large amount of available gas, there is about an order of magnitude less active star formation in the CMZ than expected \citep{LongmoreEtAl2013a,KruijssenEtAl2014,JohnstonEtAl2014}. In order to test theories of star formation, our main aim here is to measure the amount and structure of the turbulence and to determine the magnetic field. We do this for the CMZ cloud {\brick}, also known as the `Brick'.

Besides constraining fundamental parameters of {\brick}, such as the density and mass of the cloud, we focus on determining the turbulent Mach number and driving, as well as the turbulent magnetic field component. We reconstruct the driving mode of the turbulence in {\brick} and find that it is primarily solenoidal. This is in stark contrast to spiral-arm clouds, where the turbulence seems to be significantly more compressive \citep{PadoanJonesNordlund1997,Brunt2010,PriceFederrathBrunt2011,GinsburgFederrathDarling2013}. The solenoidal driving of turbulence in {\brick} may provide a possible explanation for the unusually low efficiency of dense-core and star formation in this environment.

Recent observations with the Atacama Large Millimeter/submillimeter Array (ALMA) have revealed that {\brick} is indeed a molecular cloud with a highly complex structure governed by turbulent motions \citep{RathborneEtAl2014,RathborneEtAl2015}. These high-resolution dust and molecular line observations indicate that {\brick} is filamentary, with networks of filaments having similar complexity as in nearby spiral-arm clouds \citep{AndreEtAl2014}. So far the filamentary structure inside {\brick} has not been quantified, because pre-ALMA observations did not have sufficient resolution. Here we measure the average filament column density and width in this CMZ cloud and compare our measurements to nearby spiral-arm clouds.

\subsection{Turbulence driving?}

The observations by \citet{RathborneEtAl2014,RathborneEtAl2015} demonstrate that {\brick} is highly turbulent, but it has been unclear what drives this turbulence \citep[for a discussion of potential drivers of turbulence in the CMZ, see \S5.2 in][]{KruijssenEtAl2014}. Numerical simulations have shown that turbulence decays quickly in about a crossing time \citep{ScaloPumphrey1982,MacLowEtAl1998,StoneOstrikerGammie1998,MacLow1999}. The fact that we see turbulence thus leads us to conclude that it must be driven by some physical stirring mechanism. In general, potential driving mechanisms include supernova explosions and expanding radiation fronts and shells induced by high-mass stellar feedback \citep{McKee1989,KrumholzMatznerMcKee2006,BalsaraEtAl2004,BreitschwerdtEtAl2009,PetersEtAl2011,GoldbaumEtAl2011,LeeMurrayRahman2012}, winds \citep{ArceEtAl2011}, gravitational collapse and accretion of material \citep{VazquezCantoLizano1998,KlessenHennebelle2010,ElmegreenBurkert2010,VazquezSemadeniEtAl2010,FederrathSurSchleicherBanerjeeKlessen2011,RobertsonGoldreich2012,LeeChangMurray2015}, and Galactic spiral-arm compressions of H\textsc{I} clouds turning them into molecular clouds \citep{DobbsBonnell2008,DobbsEtAl2008}, as well as magneto-rotational instability (MRI) and shear \citep{PiontekOstriker2007,TamburroEtAl2009}. Jets and outflows from young stars and their accretion disks have also been suggested to drive turbulence \citep{NormanSilk1980,MatznerMcKee2000,BanerjeeKlessenFendt2007,NakamuraLi2008,CunninghamEtAl2009,CarrollFrankBlackman2010,WangEtAl2010,CunninghamEtAl2011,PlunkettEtAl2013,PlunkettEtAl2015,OffnerArce2014,FederrathEtAl2014}. While different drivers may play a role in different environments (such as in spiral-arm clouds), \citet{KruijssenEtAl2014} found that most of these drivers are not sufficient to explain the turbulent velocity dispersions in the CMZ.

Importantly, most of these turbulence drivers primarily compress the gas (e.g., supernova explosions, high-mass stellar feedback, winds, gravitational contraction, and spiral-arm shocks), but others can directly excite solenoidal motions (e.g., MRI, jets/outflows, and shear). Our goal here is to determine the fraction of solenoidal and compressive modes in the driving of the turbulence in {\brick}. This relative fraction of driving modes is determined by the \emph{turbulence driving parameter} $b$, which is proportional to the ratio of density to velocity fluctuations, $b\propto\sigma_\rho/\sigma_v$, in a supersonically turbulent cloud \citep{FederrathKlessenSchmidt2008,FederrathDuvalKlessenSchmidtMacLow2010}. \citet{FederrathKlessenSchmidt2008} showed that purely solenoidal (rotational or divergence-free) driving corresponds to $b=1/3$, while purely compressive (potential or curl-free) driving results in $b=1$. Increasing the fraction of compressive modes in the turbulence driving from zero to unity leads to a smoothly increasing driving parameter $b$ \citep[see Fig.~8 in][]{FederrathDuvalKlessenSchmidtMacLow2010}.\footnote{Note that even if the turbulence \emph{driving field} is fully compressive ($b=1$), there is still a substantial fraction of solenoidal modes that will be excited in the \emph{velocity field} via non-linear interactions \citep{Vishniac1994,SunTakayama2003,KritsukEtAl2007,FederrathDuvalKlessenSchmidtMacLow2010}, baroclinic instability \citep{DelSordoBrandenburg2011,PadoanEtAl2016,PanEtAl2016}, and by viscosity across density gradients \citep{MeeBrandenburg2006,FederrathEtAl2011PRL}.}

Here we determine the turbulence driving parameter $b$ by measuring the standard deviation of the density fluctuations $\sigrho$ and the standard deviation of the probability distribution function (PDF) of the turbulent velocity field in {\brick}. We find that the turbulence driving in {\brick} is dominated by solenoidal shearing motions ($b<0.4$), while spiral-arm clouds have a substantial compressive driving component, $b>0.4$. Our results support the idea that shear is a typical driving mode of the turbulence in the CMZ and possibly in the centers of other galaxies, as proposed by \citet{KrumholzKruijssen2015} and Kruijssen et al., in preparation. This solenoidal driving mode can suppress star formation \citep{FederrathKlessen2012,PadoanEtAl2014} and may thus provide a possible explanation for the low SFR in the CMZ.

\subsection{Universal filament properties?}

Interstellar filaments are considered to be fundamental building blocks of molecular clouds, playing a crucial role in star formation \citep{SchneiderElmegreen1979,BalsaraEtAl2001,AndreEtAl2014}. Indeed, star-forming cores in nearby spiral-arm clouds are often located along dense filaments \citep{PolychroniEtAl2013,KonyvesEtAl2015} and young star clusters tend to form at their intersections \citep{Myers2011,SchneiderEtAl2012}. Recent observations and simulations of spiral-arm clouds show that filaments have coherent velocities \citep{HacarEtAl2013,MoeckelBurkert2015,HacarEtAl2016,SmithEtAl2016} and orientations preferentially (but not always) perpendicular to the magnetic field \citep{SugitaniEtAl2011,GaenslerEtAl2011,PalmeirimEtAl2013,Hennebelle2013,Tomisaka2014,ZhangEtAl2014,PlanckMagneticFilaments2014,PlanckMagneticFilaments2015a,PlanckMagneticFilaments2015b,PillaiEtAl2015,SeifriedWalch2015}. Most importantly, filaments seem to have a nearly universal width $W_\mathrm{fil}\sim0.1\,\pc$ \citep{ArzoumanianEtAl2011,JuvelaEtAl2012a,PalmeirimEtAl2013,MalinenEtAl2012,BenedettiniEtAl2015,KirkEtAl2015,WangEtAl2015,RoyEtAl2015,SaljiEtAl2015,KainulainenEtAl2016}.\footnote{Note that \citet{JuvelaEtAl2012a} and \citet{SaljiEtAl2015} found maximum variations of $W_\mathrm{fil}$ by a factor of $28$, while \citet{ArzoumanianEtAl2011} found maximum variations up to a factor of $10$. Thus, the term `universal' means in this context that $W_\mathrm{fil}$ definitely varies by less than two orders of magnitude, but more likely within factors of only a few around $0.1\,\pc$. Also note that \citet{SmithGloverKlessen2014} found somewhat larger values and variations of $W_\mathrm{fil}$ from simulations, in contrast to the observations in \citet{ArzoumanianEtAl2011}.} \citet{Federrath2016} provided a turbulence-regulated model for $W_\mathrm{fil}$, which is based on the sonic scale of the turbulence.

Here we show that over-dense regions are located along filaments also in the CMZ cloud {\brick}, but the average filament column density is about 1--2 orders of magnitude higher compared to nearby clouds. Surprisingly though, the average filament width is similar in {\brick} to solar neighborhood clouds. Given the significant difference in gas temperature and magnetic fields in the CMZ, it seems surprising that $W_\mathrm{fil}$ is similar in {\brick} to nearby clouds. We explain the universal value for $W_\mathrm{fil}$ with the \emph{sonic scale}---the transition scale from supersonic to subsonic turbulence, following the theoretical model developed in \citet{Federrath2016}. We find excellent agreement between the measured filament width and the predicted sonic scale, both in {\brick} and in nearby clouds.

The paper is organized as follows. Section~\ref{sec:observations} summarizes the observational data. In Section~\ref{sec:dens}, we identify filaments, measure their width and column density, and reconstruct the volume density dispersion of {\brick}. We measure the velocity PDFs of the total and turbulent (gradient-subtracted) velocity field in Section~\ref{sec:vels}. Numerical simulations to constrain the turbulent magnetic field are presented in Section~\ref{sec:mag}. We summarize all our measured and derived physical parameters of {\brick} in Table~\ref{tab:brick} of Section~\ref{sec:physics}. Sections~\ref{sec:sonicscale} and~\ref{sec:driving} provide a detailed discussion of derived sonic scale and turbulence driving parameter with comparisons to nearby clouds. A discussion of the limitations of this work are presented in Section~\ref{sec:caveats}. Our conclusions are summarized in Section~\ref{sec:conclusions}.

\section{Observational data} \label{sec:observations}

\citet{RathborneEtAl2014,RathborneEtAl2015} obtained a $1'\times3'$ mosaic of the 3~mm (90~GHz) dust continuum and molecular line emission across {\brick}, using 25~antennas as part of ALMA's Early Science Cycle~0. The interferometer used projected baselines in the range $13$--$455\,\mathrm{m}$. The correlator was configured to use four spectral windows in dual polarization mode centered at 87.2, 89.1, 99.1, and 101.1~GHz, each with 1875~MHz bandwidth and 488~kHz \mbox{($1.4$--$1.7\,\km\,\s^{-1}$)} velocity channel spacing. The {\brick} cloud was imaged on six occasions between July 29 and August 1 in 2012. Each data set was independently calibrated before being merged. All data reduction was performed using CASA \citep{casa} and Miriad \citep{Miriad}.

\subsection{Dust emission and column density derivation}
The ALMA dust continuum data were complemented with single-dish data from the \emph{Herschel} space observatory to recover the large-scale component of the dust emission. These dust emission data were then converted to gas column densities with the techniques and assumptions explained in detail in \citet{RathborneEtAl2014}. The final column density image has a pixel size of $0.35''$, an angular resolution of $1.7''$ ($\mathrm{FWHW}\sim0.07\,\pc$), and a $10\,\sigma$ sensitivity of $\sim0.25\,\mathrm{mJy}\,\mathrm{beam}^{-1} \sim4.8\times10^{22}\,\cm^{-2}$ \citep{RathborneEtAl2014}. In all the following measurements and derivations, we only use data within the $5\times10^{22}\,\cm^{-2}$ column density contour level ($\mathrm{S/N}>10$).

In addition to the combined ALMA+\emph{Herschel} column density map from \citet{RathborneEtAl2014}, we also use the \emph{Herschel}-only column density map published in \citet{LongmoreEtAl2012}. The resolution of the \emph{Herschel} data is $5''$--$36''$. The \emph{Herschel} column densities were derived based on a fit to the spectral energy distribution using five photometric bands (70, 160, 250, 350, and 500$\,\mu$m) from \emph{Herschel} Hi-GAL \citep{MolinariEtAl2010,MolinariEtAl2011}. The absolute column density level is thus better calibrated in the \emph{Herschel} map than in the ALMA+\emph{Herschel} map (see Sec.~\ref{sec:caveats}). In order to derive the average column density and mass of {\brick} we make use of the pure \emph{Herschel} map, while the ALMA+\emph{Herschel} map is used to identify filaments and to measure the column-density and volume-density dispersions.

\subsection{HNCO line data to derive gas kinematics} \label{sec:hnco}
Because the 90~GHz spectrum is rich in molecular lines, \citet{RathborneEtAl2014,RathborneEtAl2015} also obtained data cubes from 17~different molecular species in {\brick}. Combined, they provide information on the gas kinematics and chemistry within the cloud. \citet{RathborneEtAl2015} analyzed each molecular line map in detail and found that the best available overall correlation between the dust continuum and the integrated line emission are obtained with HNCO, H$_2$CS, and NH$_2$CHO. While the latter two only cover a small fraction of the cloud because of insufficient signal-to-noise (S/N), the HNCO line provides good coverage and high S/N of the dense gas above $5\times10^{22}\,\cm^{-2}$. The HNCO line brightness sensitivity is $\sim1\,\mathrm{mJy}\,\mathrm{beam}^{-1}$ per $3.4\,\km\,\s^{-1}$ channel. As discussed in \citet{RathborneEtAl2015}, HNCO has a strong dipole moment and a high excitation energy, making HNCO less susceptible to optical depth effects. We thus focus here on using HNCO to trace the global, large-scale kinematics of {\brick}. However, we emphasize that the local correlation with the dust emission is not sufficient to trace the kinematics of individual column density features on small scales. This would require data from a better (or a combination of) molecular line tracer(s), because each molecular transition can only trace certain (local) conditions of the gas. Caveats and limitations of these data are discussed in Section~\ref{sec:caveats}.

\section{Results} \label{sec:results}

\subsection{Density structure} \label{sec:dens}

Here we determine the turbulent, filamentary structure of {\brick}. We measure the characteristic width of the filaments and determine the global turbulent density fluctuations. Both the filament width and the standard deviation of the density PDF are key measurements to understand the star formation activity of {\brick}.

\begin{figure*}
\centerline{\includegraphics[width=1.0\linewidth]{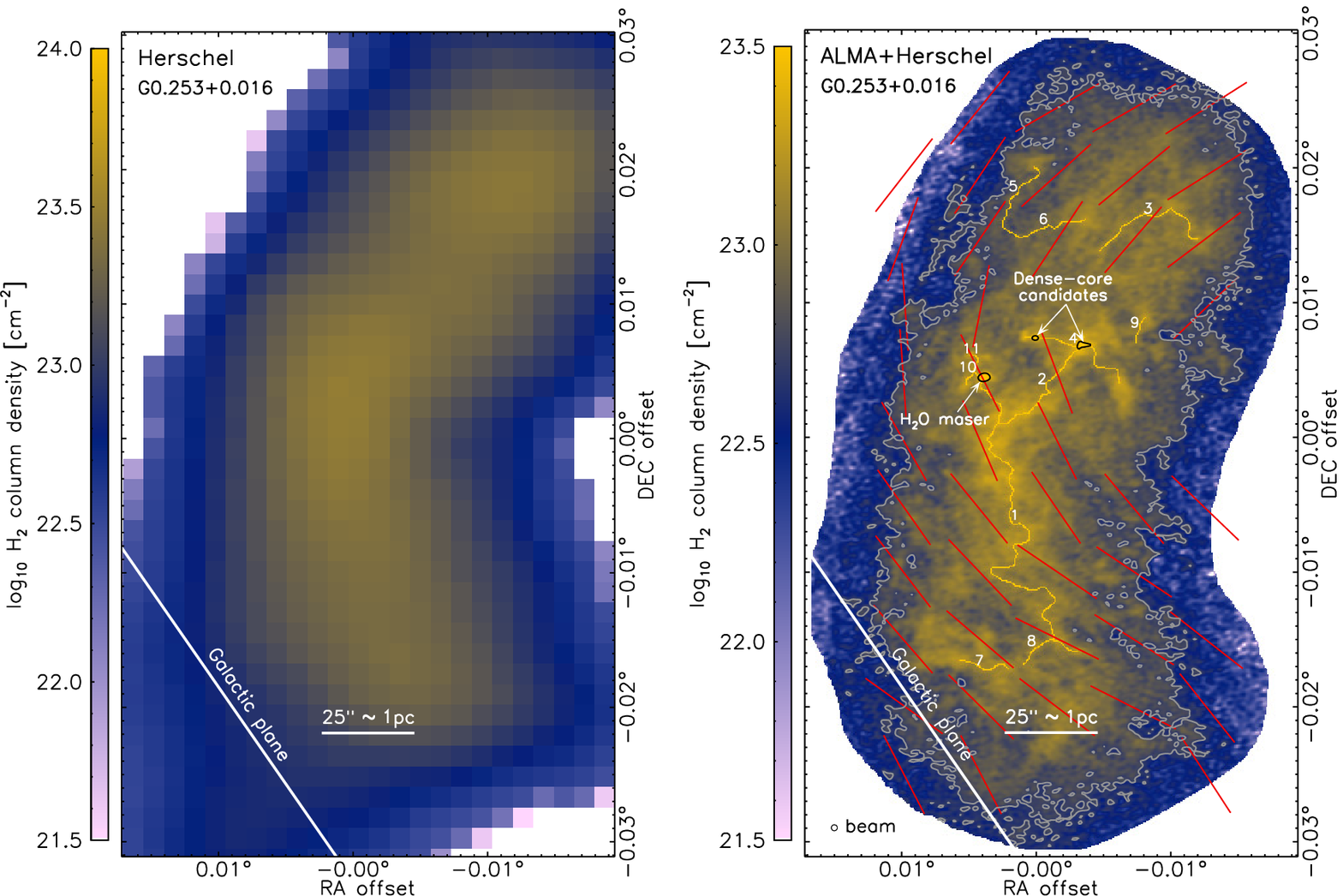}}
\caption{H$_2$ column density maps of {\brick} from \emph{Herschel} \citep{LongmoreEtAl2012} (left-hand panel) and ALMA+\emph{Herschel} \citep{RathborneEtAl2014} (right-hand panel). The \emph{Herschel} map traces the large-scale structure well, while the combined ALMA+\emph{Herschel} map reveals the internal structure of {\brick}. Using the DisPerSE algorithm we identify 11 filaments in the ALMA+\emph{Herschel} map, which are highlighted and labelled by artificially increasing the column density by a factor of 5 for each pixel belonging to a filament. The gray contour encloses gas with a column density $N\geq5\times10^{22}\,\cm^{-2}$ (lower column densities have relatively low S/N). The position of a water maser is labelled and is located along filament~1, where $N\geq2.5\times10^{23}\,\cm^{-2}$ (black contours). Another two over-dense regions (`dense-core candidates') above the same threshold stand out along filaments~2 and 4. Red lines indicate the direction of the large-scale magnetic field from polarization measurements obtained in \citet{DotsonEtAl2010}; see \citet{PillaiEtAl2015}. Both images are in equatorial coordinates: the (0,0) offset position in right ascension (RA)~and declination (DEC) is 17:46:09.59, $-$28:42:34.2~J2000.}
\label{fig:coldensimage}
\end{figure*}

First we start with the basic column density structure. Figure~\ref{fig:coldensimage} shows a side-by-side comparison of the column density maps of {\brick} from \emph{Herschel} \citep{LongmoreEtAl2012} and ALMA+\emph{Herschel} \citep{RathborneEtAl2014}, showing the substantial improvement in resolution provided by ALMA. We see a complex network of intersecting filaments in the ALMA+\emph{Herschel} map. These filaments were identified with the DisPerSE algorithm \citep{Sousbie2011,SousbieEtAl2011}, which decomposes the map into a set of persistent maxima and saddle points, which are connected to build the filament structure shown.\footnote{The filaments are identified in the column density map, i.e., they represent projected structures along the line of sight (LOS). A separation of these structures in position-position-velocity space is currently not possible with the data at hand (see discussion in Section~\ref{sec:caveats}), so we restrict ourselves to the analysis of the projected filaments. Thus individual filaments in the map may actually consist of multiple sub-filaments along the LOS, but simulations have shown that the average width of these projected filaments agrees with the average width of the intrinsic 3-dimensional filaments to within a factor of 2 (Sec.~\ref{sec:caveats}).} Note that the most important parameter in the DisPerSE algorithm is the persistence threshold, which we have set here to $5\times10^{22}\,\cm^{-2}$, i.e., $10\,\sigma$ of the sensitivity threshold of the observations (see Sec.~\ref{sec:observations}), in order to find only the most significant and dense filaments.\footnote{All other DisPerSE parameters were set to the recommended standard values. The full DisPerSE command lines used were: \texttt{mse map.fits -noTags -upSkl -periodicity 0 -cut 4.75e22 -robustness} and \texttt{skelconv map.fits.up.NDskl -noTags -toFITS -breakdown -smooth 6 -trimBelow robustness 4.75e22 -assemble 70}. We further enforced a minimum number of 5 pixels per filament.} We have experimented with higher and lower persistence thresholds and found similar filaments and similar filament column densities and widths. Decreasing or increasing the threshold by a factor of two neither significantly affects the number of identified filaments nor their average properties.\footnote{A systematic analysis of varying the persistence threshold is performed in \citet{Federrath2016}, showing that the average filament width does not significantly depend on the choice of persistence threshold, while the average column density of the filaments decreases with decreasing threshold, as expected.}

The black contours In Figure~\ref{fig:coldensimage} highlight three prominent over-dense regions with $N\geq2.5\times10^{23}\,\cm^{-2}$ \citep[one potentially active region of star formation as indicated in the map and traced by a water maser; see][]{LisEtAl1994,BreenEllingsen2011,MillsEtAl2015}. \citet{RathborneEtAl2014} used a $2\times$ higher threshold ($N>5\times10^{23}\,\cm^{-2}$) based on the fact that the column-density PDF starts to deviate from a log-normal PDF at this column-density threshold. Using $N>5\times10^{23}\,\cm^{-2}$ only selects the water-maser location, which \citet{RathborneEtAl2014} confirmed to be a coherent and bound core. Here we reduce the threshold by a factor of 2, which yields another two dense structures that we call `dense-core candidates'. We cannot confirm them as coherent structures in velocity space at this point (see discussion about the correlation of dust and molecular line emission in Sec.~\ref{sec:caveats}). However, given the uncertainties in the column-density calibration (see Sec.~\ref{sec:caveats}), the $N\geq2.5\times10^{23}\,\cm^{-2}$ threshold used here is still consistent with the deviation point in the column-density PDF from a log-normal distribution to a high column-density tail found in \citet{RathborneEtAl2014}.

The average effective diameter of the water maser and the two dense-core candidates is $0.09\,\pc$ with a variation by about a factor of 2. The filling fraction of these dense structures is only $0.0011\pm0.0001$ of the total area of {\brick}, indicating inefficient dense-core and star formation \citep[see also][]{KauffmannPillaiZhang2013}. The water maser and the two dense-core candidates are located along filaments 1, 2 and 4. Dense cores are often associated with filaments and their intersections, which is also seen in clouds in the spiral arms of the Milky Way \citep{SchneiderEtAl2012,PolychroniEtAl2013,KonyvesEtAl2015}. This suggests that filaments may be fundamental building blocks of molecular clouds, irrespective of whether the clouds are located along spiral arms or near the Galactic Center.

Finally, the red lines in the right-hand panel of Figure~\ref{fig:coldensimage} indicate the projected large-scale magnetic field direction ($B_0$) inferred from polarization measurements by \citet{DotsonEtAl2010} and further analysed in \citet{PillaiEtAl2015}. We see that some filaments are mainly parallel to $B_0$ (e.g., filaments~1 and 5), while others are primarily perpendicular to $B_0$ (e.g., filaments~2 and 4). We do not find that the filaments have a preferred orientation with respect to the large-scale magnetic field. In the following, we determine the width of these filaments and distinguish filaments primarily parallel or perpendicular to $B_0$, in order to test whether the width depends on the filament orientation.

\subsubsection{Filament profiles} \label{sec:filprof}

\begin{figure}
\centerline{\includegraphics[width=1.0\linewidth]{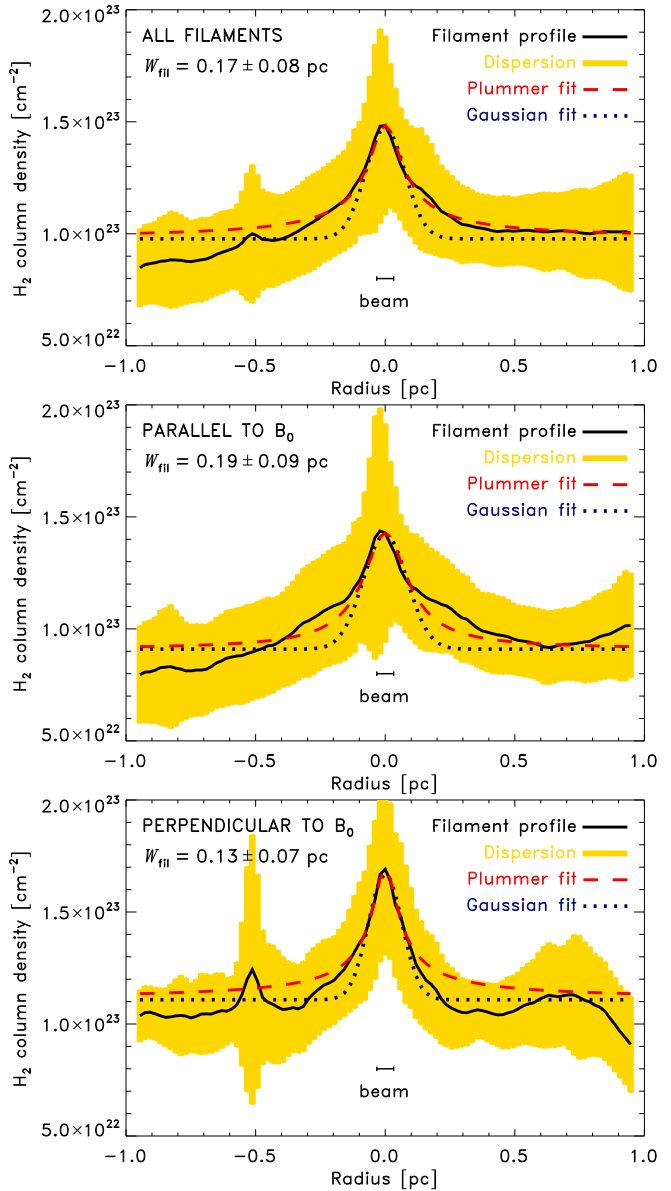}}
\caption{Top panel: Average radial profile of all the {\brick} filaments in Figure~\ref{fig:coldensimage}. Middle panel: same as top panel, but only for the filaments that are primarily parallel to the large-scale magnetic field ($B_0$). Bottom panel: same as top panel, but only for the filaments that are primarily perpendicular to $B_0$. In all panels, the shaded region shows the $1\sigma$-dispersion about the average profile. Plummer fits with Equation~(\ref{eq:plummer}) and a Gaussian fits with Equation~(\ref{eq:gauss}) are shown as dashed and dotted lines, respectively. The beam size is shown as a ruler. Both Gaussian and Plummer fits yield consistent beam-corrected filament widths of $W_\mathrm{fil}=0.17\pm0.08\,\pc$ for all filaments, $W_\mathrm{fil}=0.19\pm0.09\,\pc$ for the filaments primarily parallel to $B_0$, and $W_\mathrm{fil}=0.13\pm0.07\,\pc$ for the filaments mainly perpendicular to $B_0$.}
\label{fig:filprof}
\end{figure}

In order to measure the characteristic width of the filaments in {\brick}, we construct radial profiles centered on each individual filament in Figure~\ref{fig:coldensimage}. The procedure is similar to that applied in previous studies \citep{ArzoumanianEtAl2011,Federrath2016}. The radial profiles are computed by selecting all pixels belonging to a filament and then tracing the column density cells at a perpendicular distance $r$ to the filament as in previous studies \citep[e.g.,][and references therein]{Federrath2016}. Binning the average column density and column density dispersion in the radial distance $r$ from each filament yields the filament profile.

Figure~\ref{fig:filprof} shows the filament profile of {\brick} (the black line is the average profile and the shaded region shows the $1\sigma$-dispersion). In order to determine the filament width $W_\mathrm{fil}$, we apply two fits, one with a Gaussian profile, the other with a Plummer profile.

The Gaussian filament profile (shown as the dotted line in Figure~\ref{fig:filprof}) is defined as
\begin{equation}
\Sigma(r) = \Sigma(0) \exp{\left(-\frac{r^2}{2\sigma_\mathrm{Gauss}^2}\right)} + \Sigma_\mathrm{offset},
\label{eq:gauss}
\end{equation}
with the fit parameters $\Sigma(0)$ and $\sigma_\mathrm{Gauss}$. The filament width $W_\mathrm{fil} = 2\sqrt{2\ln 2}\,\sigma_\mathrm{Gauss}\approx2.355\,\sigma_\mathrm{Gauss}$ is defined as the FWHM of the Gaussian. The constant column density offset $\Sigma_\mathrm{offset}=1\times10^{23}\,\cm^{-2}$ was chosen to be consistent with the average column density inside the $5\times10^{22}\,\cm^{-2}$ contour of {\brick}, providing high S/N column density values.

The Plummer filament profile (shown as the dashed line in Figure~\ref{fig:filprof}) is defined as
\begin{equation}
\Sigma(r) = \Sigma(0) \left[ 1 + \left(r/\rflat\right)^2 \right]^{(1-p)/2} + \Sigma_\mathrm{offset},
\label{eq:plummer}
\end{equation}
with the fit parameters $\Sigma(0)$, $p$ and $\rflat$, where the latter is related to the filament width $W_\mathrm{fil}\approx3\,\rflat$ for $p=2$ \citep{ArzoumanianEtAl2011}. \citet{ArzoumanianEtAl2011}, \citet{ContrerasRathborneGaray2013}, \citet{SmithGloverKlessen2014} and \citet{Federrath2016} experimented with the power $p$ and found that the best fits to the filament profiles were obtained with $p\approx2$. Here we find $p=2.1\pm1.0$ for {\brick} from the Plummer fit shown in Figure~\ref{fig:filprof}.

\citet{Federrath2016} provided a theoretical model for $p=2$, which is given by the radial dependence of the density profile of two colliding planar shocks forming a filament at their intersection. In contrast to this dynamical, turbulence-regulated model for filament formation by \citet{Federrath2016}, \citet{Ostriker1964} studied the case in which the filaments are in hydrostatic equilibrium, which gives significantly steeper profiles, $p=4$, ruled out by our observations of {\brick} and previously ruled out for nearby clouds \citep{ArzoumanianEtAl2011}.\footnote{A number of previous studies find some variations in the filament-profile exponent $p$ for different clouds. \citet{NutterEtAl2008} find $p\sim3$ for the Taurus molecular cloud, \citet{PinedaEtAl2011} find $p\sim4$ for B5 in Perseus, \citet{HacarTafalla2011} find \mbox{$p=2.7$--$3.4$} for 4 filaments in L1517, \citet{ContrerasRathborneGaray2013} find that $p$ can vary between clump and inter-clump gas, and \citet{SaljiEtAl2015} find that the majority of filaments in Orion A North exhibit \mbox{$p=1.5$--$3$}, with a mode at $p=2.2$. } Other theoretical models that also produce $p=2$ are discussed in \cite{Federrath2016}, but the key difference to our turbulence-regulated model is that the other models assume (magneto)hydrostatic equilibrium and/or strongly self-gravitating filaments, which are strong assumptions. Based on our analyses of the kinematics and virial parameter of {\brick} (summarized in Tab.~\ref{tab:brick} below), we do not believe that models of hydrostatic balance or strong self-gravity represent the dynamics of the cloud well. Instead we find that {\brick} is governed by supersonic turbulence, consistent with the filament-formation model of \citet{Federrath2016}.

In order to correct for beam smearing, we performed two independent methods of beam deconvolution. First, we performed a direct Fourier-based beam deconvolution of the filament profiles. We also made an indirect deconvolution by taking the beam into account during the profile fitting. Both techniques yield consistent results. The beam-corrected filament width ($0.17\,\pc$) is $\lesssim10\%$ smaller than the beam-convoluted width ($0.18\,\pc$). Even without performing the full deconvolution, it is straightforward to see that beam smearing has a negligible effect. Taking our beam-convoluted measurement of $0.18\,\pc$ and subtracting the effect of the beam FWHM ($0.07\,\pc$; see Sec.~\ref{sec:observations}), we find the de-convolved filament width of $[(0.18\,\pc)^2-(0.07\,\pc)^2]^{1/2}=0.17\,\pc$, in excellent agreement with the direct deconvolution.

Both Gaussian and Plummer fits in Figure~\ref{fig:filprof} yield a consistent filament width of $W_\mathrm{fil}=0.17\pm0.08\,\pc$, taking into account all the filaments identified in {\brick}, where the uncertainty is estimated based on numerical simulations by \citet{SmithGloverKlessen2014}, showing that the average intrinsic 3D filament width can be up to 50\% smaller than the average projected (2D) filament width due to line-of-sight blending (see Sec.~\ref{sec:caveats}). While Figure~\ref{fig:filprof} shows the average profile, we have also fitted each of the 11 individual filaments identified in Figure~\ref{fig:coldensimage}. The distribution of the individual filament widths has a mean value of $0.18\,\pc$ and a standard deviation of $0.04\,\pc$, consistent with the fit to the average profile. The overall uncertainty of $0.08\,\pc$ thus exceeds the filament-to-filament variations by a factor of 2.

The middle and bottom panels of Figure~\ref{fig:filprof} respectively show the average profile of filaments that are primarily parallel or perpendicular to the large-scale magnetic field ($B_0$). Figure~\ref{fig:coldensimage} showed that there is no preferred orientation of the filaments with respect to $B_0$, but we can broadly classify filaments~1, 5, 6, 7 as primarily parallel to $B_0$ and filaments~2, 4, 8, 11 as primarily perpendicular to $B_0$. The other filaments are either entirely in between these limiting cases or have some sections that are parallel and other sections that are perpendicular to $B_0$. We exclude these in-between cases from the orientation analysis, but note that we have also tested to include them and did not find a significant difference in the resulting $W_\mathrm{fil}$. We obtain $W_\mathrm{fil}=0.19\pm0.09\,\pc$ for filaments primarily parallel to $B_0$, and $W_\mathrm{fil}=0.13\pm0.07\,\pc$ for filaments mainly perpendicular to $B_0$. We thus see a weak, but statistically inconclusive trend that filaments parallel to $B_0$ may be somewhat wider than filaments perpendicular to $B_0$. Such a trend may be theoretically expected, if filaments parallel to $B_0$ were created by gas flows perpendicular to $B_0$, because these flows are more impeded by the magnetic pressure of the large-scale ordered magnetic field component. By contrast, filaments perpendicular to $B_0$ are only affected by the turbulent magnetic field component ($B_\mathrm{turb}$). The formal standard deviations of the filament width for the two populations (filaments parallel or perpendicular to $B_0$) are $0.04\,\pc$ for filaments parallel and $0.06\,\pc$ for filaments perpendicular to the ordered field. Thus, even if we only consider the formal standard deviations (without taking into account the uncertainties of projection effects; see Section~\ref{sec:fil_proj_effects}), the difference in filament widths between parallel and perpendicular filaments is still insignificant. We conclude that there is no significant difference in the filament width between filaments primarily parallel or perpendicular to the ordered magnetic field.

In summary, we find that our measured filament width of $W_\mathrm{fil}=0.17\pm0.08\,\pc$ for {\brick} is somewhat wider, but still consistent within the uncertainties with $W_\mathrm{fil}$ found in clouds in the solar neighborhood, which show a characteristic width of $\mbox{$0.05$--$0.15\,\pc$}$ \citep{ArzoumanianEtAl2011,BenedettiniEtAl2015,KainulainenEtAl2016}. We provide a theoretical explanation for this in Section~\ref{sec:sonicscale}.

Figure~\ref{fig:filprof} further shows that the characteristic maximum column density (\mbox{$\Sigma_\mathrm{max}\sim1.5\times10^{23}\,\cm^{-2}$}) and the background-subtracted column density (\mbox{$\Sigma_\mathrm{max}-\Sigma_\mathrm{offset}\sim0.5\times10^{23}\,\cm^{-2}$}) of the filaments in the CMZ cloud are more than an order of magnitude higher than in nearby spiral-arm clouds \mbox{($\sim\!0.1$--$1.5\times10^{22}\,\cm^{-2}$)}. This quantifies the extreme conditions in the CMZ, leading to at least an order of magnitude higher critical densities for star formation in the CMZ compared to spiral-arm clouds \citep{KruijssenEtAl2014,RathborneEtAl2014}.

\subsubsection{Density PDF and conversion from two-dimensional (2D) to three-dimensional (3D) density dispersion} \label{sec:pdf}

The density PDF is a key ingredient for theoretical models of the SFR and efficiency \citep{KrumholzMcKee2005,Elmegreen2008,PadoanNordlund2011,HennebelleChabrier2011,FederrathKlessen2012,Federrath2013,PadoanEtAl2014,SalimFederrathKewley2015}, for predicting bound star cluster formation \citep{Kruijssen2012}, and for the initial mass function of stars \citep{PadoanNordlund2002,HennebelleChabrier2008,HennebelleChabrier2009,HennebelleChabrier2013,ChabrierHennebelle2011,VeltchevKlessenClark2011,DonkovVeltchevKlessen2012,Hopkins2012b,Hopkins2013IMF,ChabrierEtAl2014}. It is supersonic, magnetized turbulence that determines the density PDF and, in particular, its standard deviation \citep{PadoanNordlundJones1997,FederrathKlessenSchmidt2008,PadoanNordlund2011,PriceFederrathBrunt2011,KonstandinEtAl2012ApJ,MolinaEtAl2012,BurkhartLazarian2012,NolanFederrathSutherland2015,FederrathBanerjee2015}. A high-density power-law tail can develop as a consequence of gravitational contraction of the dense cores in a cloud \citep{Klessen2000,FederrathGloverKlessenSchmidt2008,KritsukNormanWagner2011,FederrathSurSchleicherBanerjeeKlessen2011,FederrathKlessen2013,GirichidisEtAl2014}.

We do not have direct access to the 3D (volume) density from observations---only to the 2D (column) density distribution\citep{BerkhuijsenFletcher2008,KainulainenEtAl2009,LombardiAlvesLada2011,SchneiderEtAl2012,SchneiderEtAl2013,KainulainenTan2013,KainulainenFederrathHenning2013,HughesEtAl2013,BerkhuijsenFletcher2015,SchneiderEtAl2015}. However, one can estimate the 3D density dispersion and the 3D density PDF by extrapolating the 2D density information given in the plane of the sky to the 3rd dimension (along the line of sight), assuming isotropy of the clouds \citep{BruntFederrathPrice2010a,BruntFederrathPrice2010b,KainulainenFederrathHenning2014}. Here we apply the technique by \citet{BruntFederrathPrice2010a} in order to reconstruct the 3D density dispersion of {\brick}.

The column density PDF of {\brick} was analyzed in detail in \citet{RathborneEtAl2014}. They find an average column density of $N_0=(9\pm2)\times10^{22}\,\cm^{-2}$ and a logarithmic column density dispersion of $\sigma_\eta = 0.34\pm0.02$ based on a log-normal fit to the normalized column density PDF of $\eta \equiv \ln(N/N_0)$. This can be transformed to the actual column density dispersion $\sigma_{N/N_0}$ using the relation for a log-normal PDF \citep[e.g.,][]{PriceFederrathBrunt2011}, $\sigma_{N/N_0} = [\exp(\sigma_\eta^2)-1]^{1/2} = 0.35\pm0.02$. This is in agreement with the direct measurement of the column density dispersion (not using a log-normal fit) from Figure~\ref{fig:coldensimage}, which yields $\sigma_{N/N_0}=0.34$. Thus, in the following, we use $\sigma_{N/N_0}=0.35\pm0.02$.

\begin{figure}
\centerline{\includegraphics[width=1.0\linewidth]{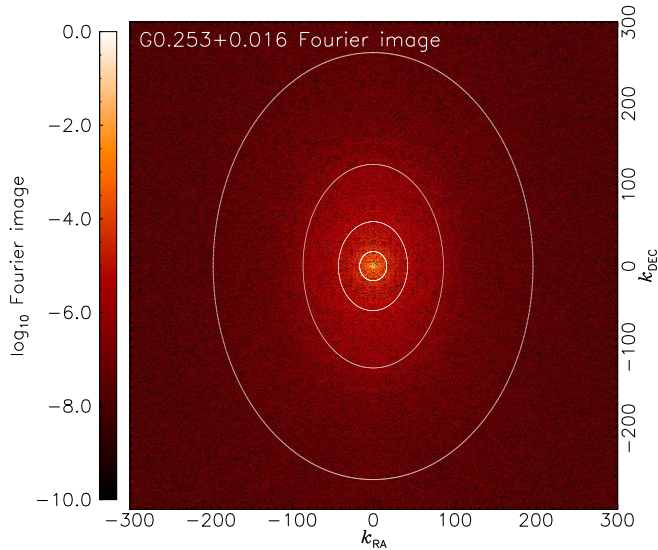}}
\caption{Fourier image of {\brick}. The intensity in the image is scaled logarithmically and normalized to the maximum intensity. Four fitted ellipses show contour levels with $10^{-4}$, $10^{-5}$, $10^{-6}$ and $10^{-7}$ of the maximum intensity. The maximum aspect ratio of the major to minor axis of the ellipses is $1.4$, which serves as a measure for the anisotropy of density structures in {\brick}, likely caused by the strong ordered magnetic field. Anisotropies of this level introduce $<\!40\%$ uncertainties in the 2D-to-3D reconstruction of the density dispersion.}
\label{fig:ftimage}
\end{figure}

In order to estimate the 3D (volume) density dispersion from $\sigma_{N/N_0}$, we use the method developed in \citet{BruntFederrathPrice2010a}. First, one measures the 2D (column) density power spectrum, $P_\mathrm{2D}(k)$ of the variable $N/N_0-1$, where $k$ is the wavenumber. Then $P_\mathrm{2D}(k)$ is multiplied by $2k$ to reconstruct the 3D (volume) density power spectrum, $P_\mathrm{3D}=2kP_\mathrm{2D}$ of the variable $\rho/\rho_0-1$ \citep[to see how well this relation between $P_\mathrm{2D}$ and $P_\mathrm{3D}$ holds for isotropic fields, we refer the reader to Figures~7 and~8 in][]{FederrathKlessen2013}. Finally, the ratio of the integrals (sums for discrete datasets) over $P_\mathrm{2D}(k)$ and $P_\mathrm{3D}(k)$ gives the density dispersion ratio
\begin{equation} \label{eq:2dto3d}
\mathcal{R}^{1/2}=\frac{\sigma_{N/N_0}}{\sigrho} = \frac{\sum_k P_\mathrm{2D}(k)}{\sum_k P_\mathrm{3D}(k)}.
\end{equation}
Note that compared to \citet{BruntFederrathPrice2010a} we here use the variable $N/N_0-1$ instead of $N/N_0$, which allows us to sum up all Fourier modes including $k=0$, while \citet{BruntFederrathPrice2010a} had to explicitly exclude the $k=0$ mode in the summation. Since subtraction of unity in our definition automatically subtracts the $k=0$ mode, the results of our and Brunt et al.'s method are identical.

\citet{BruntFederrathPrice2010a} showed that Equation~(\ref{eq:2dto3d}) holds to within 10\% for isotropic, periodic fields. They further showed that the uncertainties for non-periodic fields are somewhat higher. Here we apply mirroring of the column density map to generate a periodic dataset \citep{OssenkopfKripsStutzki2008a} in order to avoid this problem. However, Equation~(\ref{eq:2dto3d}) rests on the assumption of isotropy, so we have to check how well this assumption holds. Figure~\ref{fig:ftimage} shows the Fourier image of {\brick}. We fitted four ellipses at different intensity levels and measured the aspect ratio of their major and minor axes, in order to estimate the amount of anisotropy. The maximum major-to-minor axis ratio is $1.4$, corresponding to a moderate level of anisotropy, which is likely caused by a strong ordered magnetic field component \citep{MacLow1999,BruntFederrathPrice2010a}, observed in {\brick} \citep{PillaiEtAl2015}. Using numerical simulations, we find that for very strong magnetic guide fields that produce major-to-minor axis ratios of $2.0$, the maximum uncertainty in the 2D-to-3D reconstruction of the density dispersion is $<40\%$. Here we have a smaller axis ratio of 1.4, which is closer to typical cases of nearly isotropic fields (axes ratios up to $1.2$).\footnote{Orbital dynamics might also introduce anisotropies \citep{LongmoreEtAl2013b}, but we have not quantified this effect here.} From these considerations, we conservatively estimate the total error of our density dispersion reconstruction to be $<40\%$. Note that the uncertainty in reconstructing the full density PDF \citep{BruntFederrathPrice2010a} is higher than this, but here we only want to estimate the total 3D density dispersion and not the full 3D PDF.

Using this 2D-to-3D reconstruction technique, we find $\mathcal{R}^{1/2}=0.28\pm0.11$ for {\brick}, consistent with the average dispersion ratio of $0.27\pm0.05$ obtained from numerical simulations in \citet{FederrathDuvalKlessenSchmidtMacLow2010}. This leads to a reconstructed 3D density dispersion of $\sigrho=1.3\pm0.5$ in {\brick}. We will use $\sigrho$ in combination with an independent velocity dispersion measurement (obtained in the following section) to derive the effective driving mode of the turbulence in {\brick} in Section~\ref{sec:driving} below.

\subsection{Kinematic structure} \label{sec:vels}

\begin{figure*}
\centerline{\includegraphics[width=1.0\linewidth]{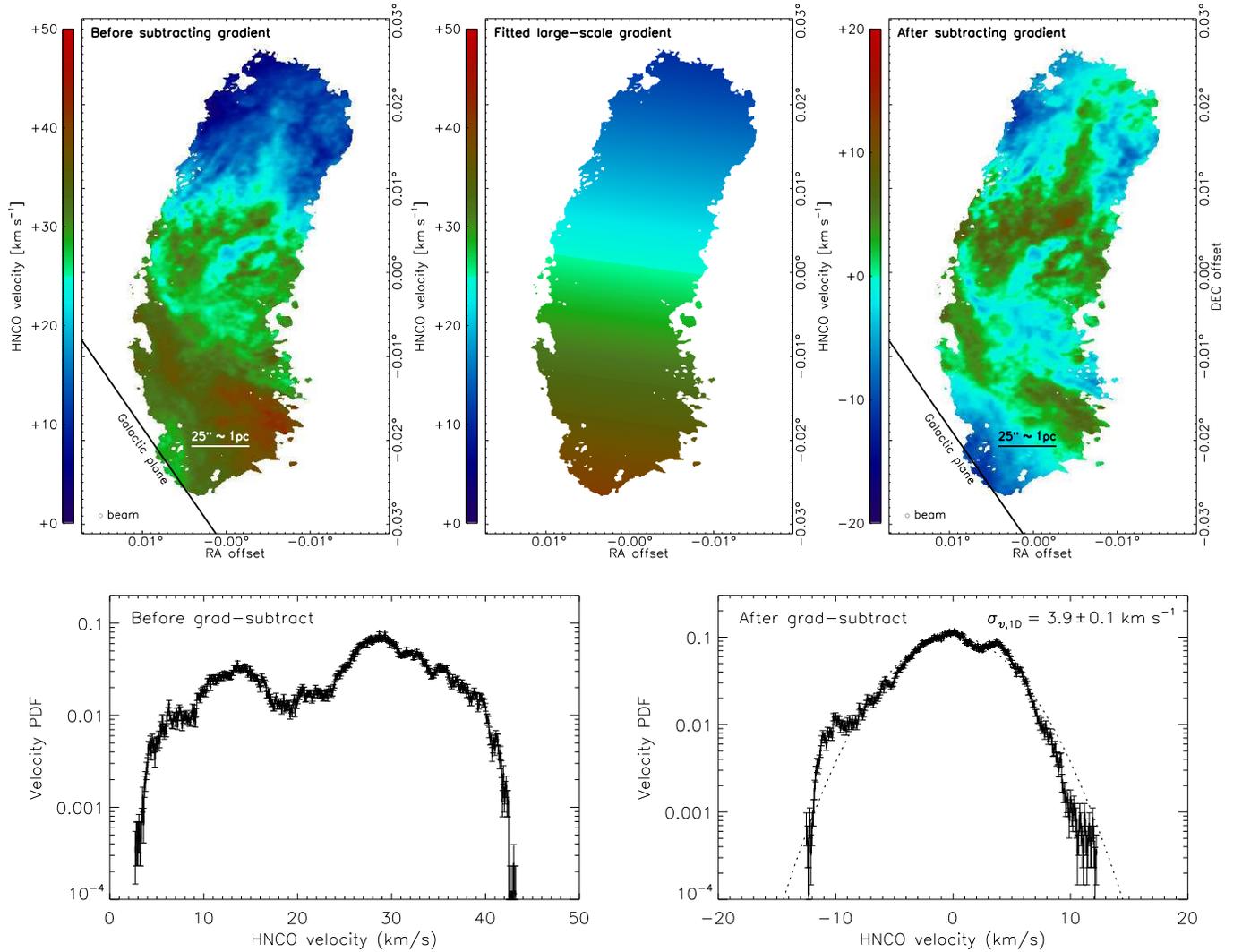}}
\caption{Top panels: maps of the HNCO intensity-weighted velocity in the {\brick}, before subtracting the large-scale velocity gradient (left-hand panel) and after subtracting it (right-hand panel). The middle panel shows the fitted gradient across {\brick}, which is likely associated with systematic motions, such as large-scale shear or rotation of the cloud. This systematic contribution must be subtracted in order to isolate the turbulent motions in the cloud. The coordinates and field of view of the maps are identical to Figure~\ref{fig:coldensimage}. Bottom panels: HNCO velocity PDF before subtracting the large-scale gradient (left-hand panel) and after subtracting it (right-hand panel). The velocity PDF after subtraction is consistent with the typical Gaussian distribution (dotted line) of a turbulent medium with a one-dimensional velocity dispersion of $\sigma_{v,\mathrm{1D}}=3.9\pm0.1\,\km\,\s^{-1}$.}
\label{fig:velocities}
\end{figure*}

Here we use the HNCO line emission of {\brick} by \citet{RathborneEtAl2015}, in order to obtain global kinematics (large-scale velocity gradient and dispersion) that we will then correlate with the global gas density dispersion obtained in the previous section. The final goal is to determine the sonic scale and the turbulence driving mode (solenoidal, mixed, or compressive). The HNCO line measurements from \citet{RathborneEtAl2015} provide the best available correlation with the ALMA dust emission and also provide the best available spatial cloud coverage, so we use it here to determine the turbulent velocity dispersion of {\brick} (cf.~Sec.~\ref{sec:hnco}).

\subsubsection{Velocity maps}
The top panels of Figure~\ref{fig:velocities} show maps of the HNCO intensity-weighted velocity (centroid velocity). The left-hand panel shows a strong and prominent velocity gradient across the long axis of {\brick}, which had been seen in earlier works \citep[e.g.,][]{RathborneEtAl2015}. This large-scale velocity gradient is likely associated with systematic rotation or shearing of the cloud. By contrast, stellar feedback or gravitational infall would produce a shock, i.e., a discontinuity, but we see a rather smooth gradient, which is most likely associated with shear (Kruijssen et al., in preparation). Such large-scale systematic motions must be subtracted in order to isolate the turbulent motions on scales smaller or equal to the size of the cloud \citep[e.g.,][]{BurkertBodenheimer2000,SurEtAl2010,FederrathSurSchleicherBanerjeeKlessen2011}. In order to isolate the turbulent motions, we fit the gradient with a plane, shown in the middle panel, and then subtract it from the original velocity map (shown in the top right-hand panel of Figure~\ref{fig:velocities}). This gradient-subtracted velocity map depicts the turbulent gas motions along the line of sight (LOS), centered on $v=0$.

\subsubsection{Velocity PDF}
Using the HNCO intensity-weighted velocity maps from the top panels of Figure~\ref{fig:velocities}, we compute velocity PDFs, shown in the bottom panels of the same Figure. The bottom left-hand panel shows the velocity PDF before subtracting the large-scale velocity gradient, while the right-hand panel shows the same after subtraction. We clearly see the two-component contributions from systematic shear or rotation of {\brick} in the PDF before subtracting the large-scale velocity gradient. The one-dimensional (1D) velocity dispersion including the systematic contribution of the gradient is $8.8\pm0.2\,\km\,\s^{-1}$, while the gradient-subtracted map yields $\sigma_{v,\mathrm{1D}}=3.9\pm0.1\,\km\,\s^{-1}$. Thus, the \emph{turbulent} velocity dispersion is significantly smaller than the total velocity dispersion.

\citet{HenshawEtAl2016} recently measured a 1D velocity dispersion of $11\,\km\,\s^{-1}$ for {\brick}, 25\% higher than our estimate that includes the contribution of the large-scale gradient. This difference arises because \citet{HenshawEtAl2016} measured the LOS velocity dispersion, while we measure the dispersion in the plane of the sky. We further correct for the large-scale gradient. However, the LOS velocity dispersion includes the contributions from the large-scale gradient and thus the dispersions and Mach numbers determined in \citet{HenshawEtAl2016} are not the purely turbulent dispersions and Mach numbers.

The gradient-subtracted PDF (bottom, right-hand panel in Figure~\ref{fig:velocities}) has the characteristic Gaussian shape of a purely turbulent medium. For example, \citet{Klessen2000} and more recently \citet{Federrath2013} show velocity PDFs from turbulence simulations and they all have this characteristic Gaussian shape. By contrast, the wide, double-peaked velocity PDF before the gradient-subtraction clearly contains non-turbulent, systematic contributions from bulk motion, shear or rotation. The Gaussian distribution in the PDF from the gradient-subtracted velocity field provides an excellent fit (shown as a dotted line), with some residual deviations. These deviations from a purely Gaussian PDF may have several sources. First, the data have intrinsic noise and measurement uncertainties. Second, the excitation conditions for the HNCO line may vary across the cloud. Third, we only subtracted the largest-scale mode (top middle panel of Figure~\ref{fig:velocities}). There might be smaller-scale modes contributing to the systematic rotation or shear, which we did not subtract. This might explain that the gradient-subtracted PDF still shows a weak second peak at a velocity $v\sim4\,\km\,\s^{-1}$ to the right of the main peak ($v=0$). Finally, turbulence has intrinsic non-Gaussian features, broadly referred to as `intermittency', leading to deviations from Gaussian statistics, especially in the tails of the PDFs \citep{FalgaronePhillips1990,PassotVazquez1998,KritsukEtAl2007,SchmidtFederrathKlessen2008,HilyBlantFalgaronePety2008,FalgaronePetyHilyBlant2009,SchmidtEtAl2009,BurkhartEtAl2009,FederrathKlessenSchmidt2009,FederrathDuvalKlessenSchmidtMacLow2010,Hopkins2013PDF,Federrath2013}.

In summary, the Gaussian fit in Figure~\ref{fig:velocities} and the standard deviation of the velocity data (without fitting) yield a consistent 1D \emph{turbulent} velocity dispersion of $\sigma_{v,\mathrm{1D}}=3.9\pm0.1\,\km\,\s^{-1}$, which we use below to derive the turbulent Mach number, the sonic scale and the turbulence driving mode of {\brick}.

\subsection{Magnetic field} \label{sec:mag}

\begin{figure}
\centerline{\includegraphics[width=1.0\linewidth]{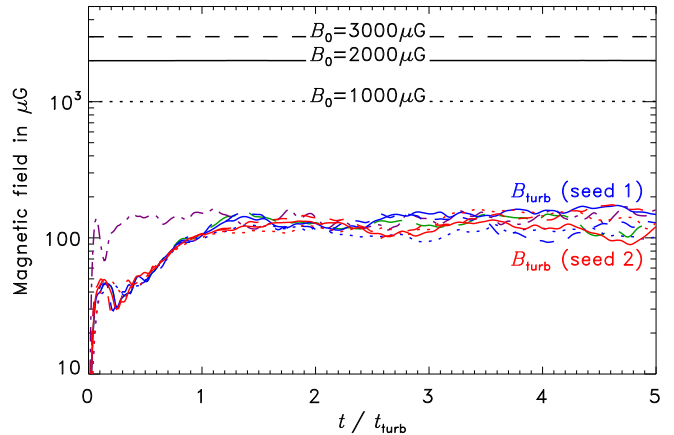}}
\caption{Magnetic field estimates for {\brick} from eight different magnetohydrodynamical turbulence simulations. Six of the eight simulations are done with three different magnetic field strengths for the ordered field component ($B_0=1000\,\mu\G$ as dotted lines, $B_0=2000\,\mu\G$ as solid lines, and $B_0=3000\,\mu\G$ as dashed lines---note that $B_0\!=\!\mathrm{const}$ because of magnetic-flux conservation), constrained by observations \citep{PillaiEtAl2015}, each one evolved with two different random seeds for the turbulence (seed 1 and 2). These six simulations were all run with solenoidal driving and a resolution of $256^3$ grid cells. Another two simulations are shown with $B_0=2000\,\mu\G$ and seed~1, but either using compressive driving (dash-dot line) or higher resolution with $512^3$ grid cells (long-dashed line). We find that the turbulent (un-ordered) field component $B_\mathrm{turb}$ can only grow to about \mbox{$100$--$200\,\mu\G$} in all cases.
}
\label{fig:magnetic}
\end{figure}

Magnetic fields play an important role for the structure of molecular clouds and for star formation \citep{PadoanEtAl2014,LiEtAl2014}. The magnetic field may be exceedingly important near the Galactic Center, where the field seems to be particularly strong \citep{TsuboiEtAl1986,YusefZadehMorris1987,SofueEtAl1987,ChussEtAl2003,Ferriere2010,CrockerEtAl2010,CrockerEtAl2011,BallyEtAl2014}. Recent measurements of the magnetic field in {\brick} find a strong ordered magnetic field component with several $\mathrm{m}\G$, roughly following the large-scale bending of the cloud \citep{PillaiEtAl2015}. Based on their measurement of the standard deviation of the residual polarization angle $\sigma_\phi=9.3\pm0.9\,\mathrm{deg}$, \citet{PillaiEtAl2015} find a total magnetic field strength of $B_\mathrm{tot}=5.4\pm0.5\,\mathrm{m}\G$ by assuming a volume number density of $n=8\times10^4\,\cm^{-3}$ from \citet{LongmoreEtAl2012}. For this, \citet{PillaiEtAl2015} use the \citet{ChandrasekharFermi1953} method, 
\begin{equation} \label{cfmethod}
B_\mathrm{tot} = f \sqrt{4\pi\rho}\; \frac{\sigma_{v,\mathrm{1D}}}{\sigma_\phi}
\end{equation}
where $f\approx0.5$, $\rho=n\mu_\mathrm{mol}m_\mathrm{H}$ is the volume density based on number density ($n$), mean molecular weight ($\mu_\mathrm{mol}$) and mass of the hydrogen atom ($m_\mathrm{H}$), and $\sigma_{v,\mathrm{1D}}$ is the one-dimensional turbulent velocity dispersion. The {\brick} velocity dispersion $\sigma_{v,\mathrm{1D}}=6.4\pm0.4\,\km\,\s^{-1}$ \citep{KauffmannPillaiZhang2013} used in \citet{PillaiEtAl2015} is consistent with our measurement from the previous subsection.\footnote{Note, however, that the $\sigma_{v,\mathrm{1D}}$ in \citet{KauffmannPillaiZhang2013} was measured inside 7 individual $\pc$-sized fragments identified in {\brick}. Assuming that the turbulence acts similarly across the cloud, the 1D velocity dispersion within individual cloud fragments might be similar to the cloud-wide velocity dispersion with the largest-scale mode subtracted (Fig.~\ref{fig:velocities}).} However, the average volume number density $n$ reported in \citet{LongmoreEtAl2012} and used in \citet{PillaiEtAl2015} is incorrect. The correct value is at least 4 times smaller. Based on the \emph{Herschel} map in Figure~\ref{fig:coldensimage}, we find $n=(1.3\pm0.7)\times10^4\,\cm^{-3}$ (see Tab.~\ref{tab:brick}). Using this corrected volume density, we adjust the \citet{PillaiEtAl2015} measurement to $B_\mathrm{tot}=2.2\pm0.9\,\mathrm{m}\G$, where we have propagated the uncertainty in $n$ into $B_\mathrm{tot}$.

The relatively small standard deviation of the residual polarization angle $\sigma_\phi$ measured in \citet{PillaiEtAl2015} means that the ordered field component $B_0$ in {\brick} is significantly larger than the turbulent (un-ordered) field component $B_\mathrm{turb}$. Note that $B_\mathrm{tot}=B_0+B_\mathrm{turb}$. \citet{PillaiEtAl2015} provide an upper limit, $B_\mathrm{turb}^2/B_0^2 \leq 0.5$, which leads to $B_\mathrm{turb} \leq B_\mathrm{tot}/5$. While their constraint already shows that $B_\mathrm{turb}$ is significantly smaller than $B_\mathrm{tot}$, \citet{PillaiEtAl2015} did not provide a direct measurement of $B_\mathrm{turb}$. The turbulent field component is important, because it determines the small-scale magnetic pressure, while $B_0$ is primarily associated with the large-scale magnetic tension in {\brick}.

Here we determine the turbulent magnetic field component $B_\mathrm{turb}$ by running magnetohydrodynamical turbulence simulations following the methods in \citet{FederrathDuvalKlessenSchmidtMacLow2010,FederrathEtAl2011PRL}. These simulations are fully determined by the turbulent velocity dispersion measured for {\brick} in the previous subsection, the driving of the turbulence (solenoidal versus compressive) and the ordered magnetic field component measured in \citet{PillaiEtAl2015}, adjusted to the correct volume density (see above). We initialize three different values for the ordered magnetic field, $B_0=1000$, $2000$, and $3000\,\mu\G$ to cover the uncertainty range in $B_0$. For each of these field strengths, we perform simulations with two different random seeds in order to estimate the statistical fluctuations in $B_\mathrm{turb}$. All simulations use a resolution of $256^3$ grid points and purely solenoidal driving of the turbulence \citep{FederrathDuvalKlessenSchmidtMacLow2010}. We also re-run one of the simulations (case $B_0=2000\,\mu\G$ with seed 1), but with a higher resolution of $512^3$ grid cells in one case and in another case with fully compressive driving. We find no significant difference in $B_\mathrm{turb}$ for either resolution or different driving of the turbulence.

Figure~\ref{fig:magnetic} shows the result of the 8 turbulence simulations (3 different $B_0$ with seed 1 and seed 2 each, one simulation with higher resolution, and another simulation with compressive driving instead of solenoidal driving). Shown are the ordered ($B_0$) and turbulent ($B_\mathrm{turb}$) magnetic field components as a function of time in units of the turbulent crossing time $t_\mathrm{turb}$. Note that the turbulence becomes fully developed during the initial transient phase, $t/t_\mathrm{turb}\lesssim1$--$2$. Once the turbulence is fully established, $B_\mathrm{turb}$ only fluctuates within \mbox{$100$--$200\,\mu\G$} in all simulations, independent of $B_0$, the driving or the resolution of the simulations. We determine the time- and simulation-averaged value and find $B_\mathrm{turb}=130\pm50\,\mu\G$, where we have assumed the same relative uncertainty as in $B_\mathrm{tot}$ from the observations, i.e., 40\%. The physical reason for our finding that $B_\mathrm{turb}$ is only about 1/10 of $B_\mathrm{tot}$ is that $B_0$ is so strong that the turbulence can hardly tangle the magnetic field on small scales to build up $B_\mathrm{turb}$.\footnote{We are currently performing a parameter study in which we systematically vary $B_0$ for fixed $\mach$, to determine the dependence of $B_\mathrm{turb}$ on $B_0$. Preliminary results suggest that $B_\mathrm{turb}$ decreases monotonically with increasing $B_0$ in the strong guide-field regime.} Our simulation results are consistent with the small standard deviation of the residual polarization angle $\sigma_\phi$ measured in {\brick} \citep{PillaiEtAl2015}.

In the following we will use the derived turbulent magnetic field component to compute the \emph{turbulent} plasma $\beta$ parameter, which is required to estimate the sonic scale, the turbulent driving, and the star formation rate of {\brick}.

\section{Physical parameters of {\brick}} \label{sec:physics}

\begin{table*}
\caption{Physical parameters of {\brick} in the CMZ.}
\label{tab:brick}
\def\arraystretch{1.3}
\setlength{\tabcolsep}{2.6pt}
\begin{tabular}{lccc}
\hline
Physical Parameter & Symbol/Definition & Mean$\,$(Standard Deviation) & Comment (Reference) \\
\hline
\emph{Measured physical parameters:} \\
Area & $A$ & $17\,(1)\,\pc^2$ & From Figs.~\ref{fig:coldensimage}, \ref{fig:velocities}; (Refs.~1) \\
H$_2$ column density & $N_0$ & $1.9\,(1.0)\times10^{23}\,\cm^{-2}$ & From Fig.~\ref{fig:coldensimage}; (Refs.~2)  \\
Filament width & $W_\mathrm{fil}$ & $0.17\,(0.08)\,\pc$ & From Figs.~\ref{fig:coldensimage}, \ref{fig:filprof} \\
2D-to-3D density dispersion ratio & $\mathcal{R}^{1/2}$ & $0.28\,(0.11)$ & From Fig.~\ref{fig:coldensimage}; Eq.~(\ref{eq:2dto3d}); (Ref. 3) \\
1D turbulent+shear velocity dispersion & $\sigma_{v,\mathrm{tot,1D}}$ & $8.8\,(0.2)\,\km\,\s^{-1}$ & From Fig.~\ref{fig:velocities}, with gradient \\
1D turbulent velocity dispersion & $\sigma_{v,\mathrm{1D}}$ & $3.9\,(0.1)\,\km\,\s^{-1}$ & From Fig.~\ref{fig:velocities}, grad.~subtracted \\
\hline
\emph{Derived from numerical simulations:} \\
Turbulent magnetic field & $B_\mathrm{turb}$ & $130\,(50)\,\mu\Gauss$ & From Fig.~\ref{fig:magnetic}; Sec.~\ref{sec:mag} \\
\hline
\emph{Taken from the literature:} \\
Log.~column density dispersion & $\sigma_\eta$ & $0.34\,(0.02)$ & $\eta=\ln(N/N_0)$; (Refs.~2) \\
Gas temperature & $T$ & $100\,(50)\,\mathrm{K}$ & (Refs.~4) \\
Dust temperature & $T_\mathrm{dust}$ & $20\,(1)\,\mathrm{K}$ & (Refs.~2) \\
Total (ordered+turbulent) magnetic field & $B_\mathrm{tot}$ & $2.2\,(0.9)\,\mathrm{m}\Gauss$ & (Ref.~5) \\
Mean molecular weight per unit $m_\mathrm{H}$ & $\mu_\mathrm{mol}$ & $2.8$ & $m_\mathrm{H}$: mass of an H atom (Ref.~6)\\
\hline
\emph{Derived physical parameters:} \\
Effective diameter & $L=2\,(A/\pi)^{1/2}$ & $4.7\,(0.1)\,\pc$ & \\
Mass$^a$ & $M=N_0\mu_\mathrm{mol}m_\mathrm{H}A$ & $7.2\,(3.8)\times10^4\,\msol$ & \\
H$_2$ volume number density$^b$ & $n_0 = N_0 / L$ & $1.3\,(0.7)\times10^4\,\cm^{-3}$ & \\
Volume density & $\rho_0 = n_0 \mu_\mathrm{mol} m_\mathrm{H}$ & $6.2\,(3.3)\times10^{-20}\,\g\,\cm^{-3}$ & \\
Column density dispersion & $\sigma_{N/N_0}=[\exp(\sigma_\eta^2)-1]^{1/2}$ & $0.35\,(0.02)$ & (Ref.~7) \\
Volume density dispersion & $\sigrho=\sigma_{N/N_0}/\mathcal{R}^{1/2}$ & $1.3\,(0.5)$ & Eq.~(\ref{eq:2dto3d}); (Ref.~3) \\
Sound speed (isothermal) & $\cs=[k_\mathrm{B}T/(\mu_\mathrm{p}m_\mathrm{H})]^{1/2}$ & $0.60\,(0.15)\,\km\,\s^{-1}$ & $\mu_\mathrm{p}=2.33$ (Ref.~6) \\
Turbulent Alfv\'{e}n speed & $\va=B_\mathrm{turb}/(4\pi \rho_0)^{1/2}$ & $1.5\,(0.7)\,\km\,\s^{-1}$ & \\
Turbulent plasma beta & $\beta=2\,\cs^2/\va^2$ & $0.34\,(0.35)$ & \\
3D turbulent+shear velocity dispersion & $\sigma_{v,\mathrm{tot,3D}}=3^{1/2}\sigma_{v,\mathrm{tot,1D}}$ & $15.2\,(0.3)\,\km\,\s^{-1}$ & \\
3D turbulent velocity dispersion & $\sigma_{v,\mathrm{3D}}=3^{1/2}\sigma_{v,\mathrm{1D}}$ & $6.8\,(0.2)\,\km\,\s^{-1}$ & \\
Virial parameter (turbulence+shear) & $\alpha_\mathrm{vir,tot}=5\sigma_{v,\mathrm{tot,3D}}^2/(\pi G L^2 \rho_0)$ & $4.3\,(2.3)$ & \\
Virial parameter (turbulence only) & $\alpha_\mathrm{vir}=5\sigma_{v,\mathrm{3D}}^2/(\pi G L^2 \rho_0)$ & $0.85\,(0.45)$ & \\
Freefall time & $t_\mathrm{ff}=[3\pi/(32G\rho_0)]^{1/2}$ & $0.27\,(0.14)\,\mathrm{Myr}$ & \\
Turbulent crossing time & $t_\mathrm{turb}=L/\sigma_{v,\mathrm{3D}}$ & $0.67\,(0.03)\,\mathrm{Myr}$ & \\
Turbulent energy dissipation rate & $\eps_\mathrm{turb}=M\sigma_{v,\mathrm{3D}}^2/(2t_\mathrm{turb})$ & $1.5\,(0.8)\times10^{36}\,\mathrm{erg}\,\s^{-1}$ & \\
3D turbulent sonic Mach number & $\mach=\sigma_{v,\mathrm{3D}}/\cs$ & $11\,(3)$ & \\
3D turbulent Alfv\'{e}n Mach number & $\macha=\sigma_{v,\mathrm{3D}}/\va$ & $4.6\,(2.1)$ & \\
Sonic scale & $\ls=L\mach^{-2}(1+\beta^{-1})$ & $0.15\,(0.11)\,\pc$ & Eq.~(\ref{eq:ls}); (Refs.~8) \\
Turbulence driving parameter & $b=\sigrho\mach^{-1}(1+\beta^{-1})^{1/2}$ & $0.22\,(0.12)$ & Eq.~(\ref{eq:b}); (Refs.~9) \\
\hline
\emph{Derived star formation parameters:} \\
Log-critical density & $s_\mathrm{crit}=\mathrm{Eq.~(\ref{eq:scrit})}$ & $2.3\,(1.2)$ & Eq.~(\ref{eq:scrit}); (Ref.~10) \\
Critical number density & $n_\mathrm{crit}=n_0\exp(s_\mathrm{crit})$ & $1.0\,(1.4)\times10^5\,\cm^{-3}$ & (Ref.~10) \\
Star formation rate per freefall time & $\sfrff=\mathrm{Eq.~(\ref{eq:sfrff})}$ & $0.042\,(0.030)$ & Eq.~(\ref{eq:sfrff}); (Ref.~10) \\
Star formation rate & $\mathrm{SFR}=\sfrff\,M/t_\mathrm{ff}$ & $1.1\,(0.8)\times10^{-2}\,\msol\,\yr^{-1}$ & Eq.~(\ref{eq:sfr}); (Ref.~10) \\
\hline
\end{tabular}
\begin{minipage}{\linewidth}
\textbf{Notes.} All physical parameters are derived for pixels that fall within the $5\times10^{22}\,\cm^{-2}$ ($10\,\sigma$ sensitivity) column density contour shown in Figure~\ref{fig:coldensimage} and where the HNCO intensity-weighted velocity has valid measurements (see Figure~\ref{fig:velocities}). This defines the fixed area $A=17\,(1)\,\pc^2$ within which we derive and report all physical parameters of {\brick}. All uncertainties were propagated based on each independent parameter. The standard deviation of each parameter is provided in brackets; we note that some of the parameters do not have Gaussian probability distributions, e.g., $\beta=0.34(0.35)$, which must not be read as $\beta$ having a finite probability to be negative (by definition it must not), instead this is a consequence of the skewed distribution of $\beta$. Nevertheless, the standard deviation is always a useful measure of the uncertainty in each parameter \citep{DAgostini2004}.
References: (1) assuming a distance of $8.3\,(0.3)\,\mathrm{k}\pc$ \citep{Malkin2013,ZhuShen2013,ReidEtAl2014}, (2) \citet{LongmoreEtAl2012}, \citet{RathborneEtAl2014}, (3) \citet{BruntFederrathPrice2010a}, (4) \citet{LisEtAl2001}, \citet{MillsMorris2013}, \citet{AoEtAl2013}, \citet{BallyEtAl2014}, \citet{GinsburgEtAl2016}, (5) \citet{PillaiEtAl2015}; note that the magnetic field measurement of $5.4\,(0.5)\,\mathrm{m}\G$ in \citet{PillaiEtAl2015} was adjusted to reflect the correct volume density $n_0=1.3\,(0.7)\times10^4\,\cm^{-3}$ of {\brick}, because the volume density reported in \citet{LongmoreEtAl2012} is incorrect. We further propagated the uncertainty of $n$ into the uncertainty of $B_\mathrm{tot}$. (6) \citet{KauffmannEtAl2008}. (7) \citet{PriceFederrathBrunt2011}. (8) \citet{FederrathKlessen2012}, \citet{Federrath2016}. (9) \citet{FederrathKlessenSchmidt2008}, \citet{FederrathDuvalKlessenSchmidtMacLow2010}, \citet{PadoanNordlund2011}, \citet{MolinaEtAl2012}, \citet{FederrathBanerjee2015}, \citet{NolanFederrathSutherland2015}. (10) \citet{FederrathKlessen2012}.
$^a$Note that the mass of $1.3\times10^5\,\msol$ derived by \citet{LongmoreEtAl2012} is a factor of 1.8 higher than our estimate, because \citet{LongmoreEtAl2012} computed the mass in an area of $1.3\times10^{5}\,\msol/(10^{23}\,\cm^{-2}\,\mu_\mathrm{mol}m_\mathrm{H})=58\,\pc^2$, which is significantly larger than what we define here for the area of {\brick}. Note that the effective radius of $2.8\,\pc$ reported in \citet{LongmoreEtAl2012} also corresponds to a significantly smaller area ($25\,\pc^2$) compared to the area used for their mass estimate. Here we derive all physical quantities consistently in a fixed area $A=17\,(1)\,\pc^2$ (see above). $^b$The average volume density of $8\times10^4\,\cm^{-3}$ reported in \citet{LongmoreEtAl2012} is incorrect because of an error in the script from which that value was derived. The corrected value derived here is $n_0=1.3\,(0.7)\times10^4\,\cm^{-3}$.
\end{minipage}
\end{table*}

Here we derive new physical parameters of {\brick} based on our measurements of the volume density dispersion, the velocity PDFs and the magnetic field simulations from the previous section. Table~\ref{tab:brick} provides a comprehensive list of all measured parameters, data taken from the literature, and the derived physical parameters. We provide the defining equation for each parameter and list the mean and standard deviation for each of them. Comments and references are provided in the last column.

We note that all of the measured and derived physical parameters were consistently determined within the $5\times10^{22}\,\cm^{-2}$ ($10\,\sigma$ sensitivity) column density contour and for all pixels that had valid HNCO intensity-weighted velocity measurements. This defines a fixed area $A=17\pm1\,\pc^2$ within which we derive and report all physical parameters of {\brick}. All uncertainties were propagated based on each independent parameter. We adopt a mean molecular weight per unit hydrogen mass of $\mu_\mathrm{mol}$ = 2.8 for a cloud of 71\% molecular hydrogen gas, 27\% helium, and 2\% metals \citep[e.g.,][]{KauffmannEtAl2008}.

A few specific points should be highlighted. First, we distinguish and list both the total (turbulent+shear) velocity dispersion and the gradient-subtracted, purely turbulent velocity dispersion. For the derivation of the sonic scale and turbulence driving parameter in {\brick} (discussed in detail below), the purely turbulent velocity dispersion is the relevant quantity. Second, the total magnetic field measurement of $5.4\,(0.5)\,\mathrm{m}\G$ in \citet{PillaiEtAl2015} was adjusted to $B_\mathrm{tot}=2.2\,(0.9)\,\mathrm{m}\Gauss$ in order to reflect the measured volume density $n_0=1.3\,(0.7)\times10^4\,\cm^{-3}$ of {\brick}. Third, the mass $M=7.2\,(3.8)\times10^4\,\msol$ of {\brick} derived here is a factor of 1.8 smaller than reported in \citet{LongmoreEtAl2012}. This is because the area used to define {\brick} in \citet{LongmoreEtAl2012} was based on \emph{Herschel} column density contours rather than the area with significant HNCO emission in the ALMA data cubes, resulting in a much larger area ($58\,\pc^2$ vs.~$17\,\pc^2$).

\section{The sonic scale and filament width} \label{sec:sonicscale}
Interstellar filaments are considered to be important building blocks of the dense star-forming phase of molecular clouds \citep{AndreEtAl2010,AndreEtAl2014}. Star formation often seems to be associated with such dense filaments and, in particular, their intersections \citep{SchneiderEtAl2012}. Here we find that {\brick} in the CMZ has similar filament properties as seen in spiral-arm clouds, e.g., that over-dense regions are located along filaments (cf.~Figure~\ref{fig:coldensimage}). It is remarkable that the filament width of \mbox{$0.05$--$0.15\,\pc$} found in observations of nearby spiral-arms clouds in the Milky Way is close to universal \citep{ArzoumanianEtAl2011,BenedettiniEtAl2015,KainulainenEtAl2016,Federrath2016}. It is even more remarkable that we find here a filament width of $W_\mathrm{fil}=0.17\pm0.08\,\pc$ for the CMZ cloud {\brick}, consistent with $W_\mathrm{fil}$ in the solar neighborhood. We can explain the similar widths of the filaments in both environments with the following theoretical model.

In our model, the filament width is determined by the sonic scale of the turbulence \citep{ArzoumanianEtAl2011,Federrath2016}. The sonic scale marks the transition from supersonic turbulence on large scales to subsonic turbulence on small scales \citep{VazquezBallesterosKlessen2003}. It is defined as \citep{FederrathKlessen2012,Federrath2016}\footnote{Note that the definition of the sonic scale in Eq.~(45) in \citet{Hopkins2013IMF} is similar to ours and yields the same dependence on $\mach$, but it does not incorporate the magnetic pressure contribution that we include here and first introduced as the 'magneto-sonic' scale in Eq.~(22) in \citet{FederrathKlessen2012}.}
\begin{equation} \label{eq:ls}
\ls = L\mach^{-2}\left(1+\beta^{-1}\right),
\end{equation}
where $L$, $\mach=\sigma_{v,\mathrm{3D}}/\cs$ and $\beta$ are the cloud scale (diameter), the 3D turbulent sonic Mach number, and the ratio of thermal to magnetic pressure, plasma $\beta=p_\mathrm{thermal}/p_\mathrm{magnetic}=2\cs^2/\va^2$ of the cloud. Inserting $L=4.7\pm0.1\,\pc$, $\mach=11\pm3$, and $\beta=0.34\pm0.35$ based on our measurements and values taken from the literature summarized in Table~\ref{tab:brick}, we find a sonic scale of
\begin{equation}
\ls=0.15\pm0.11\,\pc
\end{equation}
for {\brick}. This is in excellent agreement with the filament width that we measured for {\brick} in Figure~\ref{fig:filprof}. It supports the idea that the sonic scale of magnetized turbulence given by Equation~(\ref{eq:ls}) may control the width of interstellar filaments, not only in nearby clouds \citep{Federrath2016}, but also in the CMZ.

We have to add the caveat that Equation~(\ref{eq:ls}) is generally only applicable for clouds where the magnetic field is primarily turbulent, i.e., $B_\mathrm{turb}>B_0$. This does not seem to be the case in {\brick} (see Section~\ref{sec:mag}), so we have to perform a more careful analysis of the orientation of the filaments with respect to the large-scale ordered magnetic field component $B_0$. In Figure~\ref{fig:filprof} we found that filaments parallel to $B_0$ are somewhat wider than filaments perpendicular to $B_0$, however, this is merely a trend that is not statistically significant given the uncertainties in the measured filament width. So while Equation~(\ref{eq:ls}) only takes the turbulent magnetic pressure into account and would thus theoretically only apply to the filaments perpendicular to $B_0$, it seems to provide a good match to the measured filament widths, irrespective of the filament orientation.

\section{The effective turbulent driving} \label{sec:driving}

Theoretical and numerical studies have shown that the density fluctuations ($\sigrho$) in a turbulent medium correlate with the Mach number ($\mach$) and the driving of the turbulence, which is controlled by \emph{the turbulence driving parameter} $b$ \citep{FederrathKlessenSchmidt2008,FederrathDuvalKlessenSchmidtMacLow2010,PriceFederrathBrunt2011,KonstandinEtAl2012ApJ,NolanFederrathSutherland2015,FederrathBanerjee2015},
\begin{equation}
\sigrho = b\,\mach\left(1+\beta^{-1}\right)^{-1/2},
\label{eq:b}
\end{equation}
with
\begin{equation}
b=
\begin{cases}
1/3: \text{purely solenoidal driving} \\
0.4: \text{natural mixture} \\
1: \text{purely compressive driving}.
\end{cases}
\label{eq:bvar}
\end{equation}
Equation~(\ref{eq:b}) includes the magnetic pressure contribution through the thermal-to-magnetic pressure ratio, plasma $\beta$ \citep{PadoanNordlund2011,MolinaEtAl2012}. Note that in the absence of magnetic fields, where $\beta\to\infty$, the relation simplifies to $\sigrho=b\mach$. While Equation~(\ref{eq:bvar}) lists the three special cases (solenoidal, mixed, compressive), the driving parameter can vary continuously in the range $1/3\lesssim b \lesssim 1$. An increasing $b$ value corresponds to an increasing fraction of compressive modes in the turbulent driving mechanism. The special case called `natural mixture' is close to solenoidal driving and refers to the situation where the turbulent driving modes are randomly distributed in all three dimensions \citep[see Fig.~8 in][]{FederrathDuvalKlessenSchmidtMacLow2010}.

Given our measurements of $\sigrho=1.3\pm0.5$, $\mach=11\pm3$ and $\beta=0.34\pm0.35$ in {\brick} from the previous sections and summarized in Table~\ref{tab:brick}, we can solve Equation~(\ref{eq:b}) for the turbulence driving parameter and find
\begin{equation}
b=\sigrho\mach^{-1}(1+\beta^{-1})^{1/2}=0.22\pm0.12.
\end{equation}
This result means that turbulence in {\brick} is primarily caused by a solenoidal driving mechanism.

\subsection{The density dispersion--Mach number relation}

\begin{figure*}
\centerline{\includegraphics[width=0.72\linewidth]{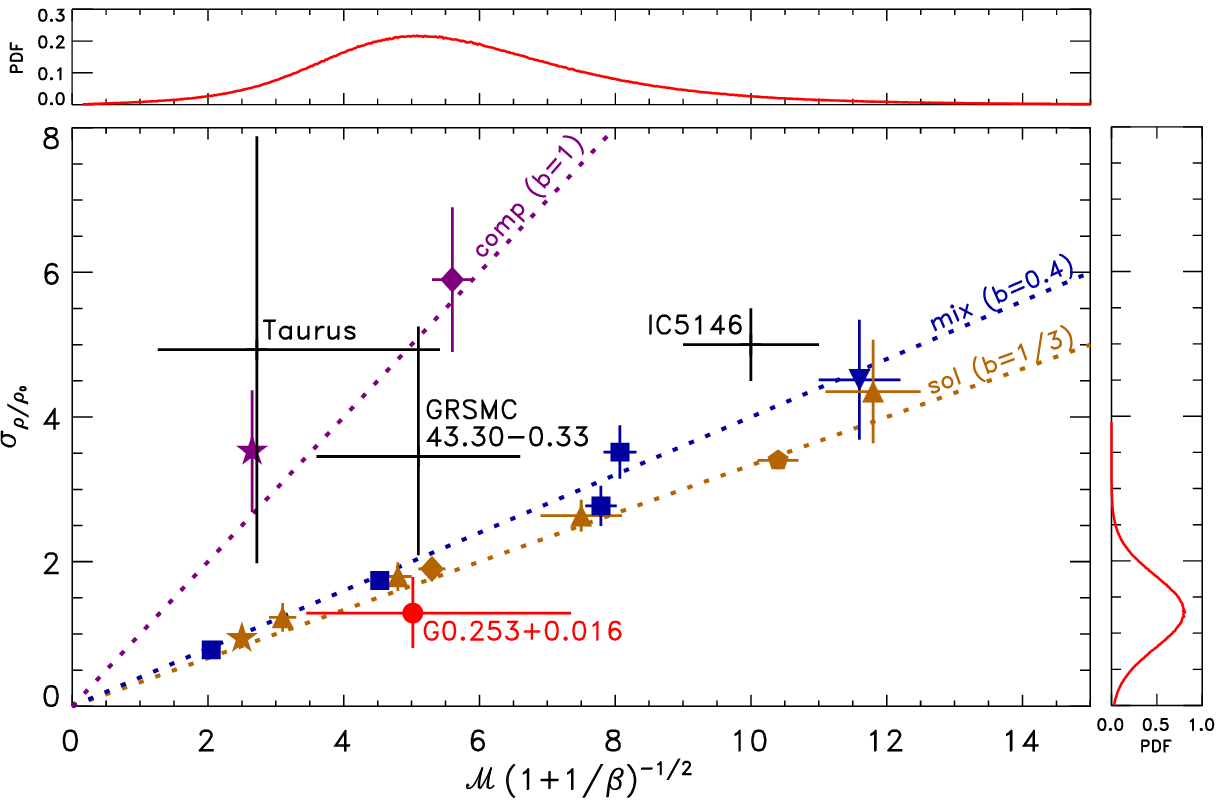}}
\caption{Relation between the turbulent density fluctuations ($\sigrho$) and the combination of sonic Mach number ($\mach$) and plasma $\beta$. This relation given by Equation~(\ref{eq:b}), defines the turbulence driving parameter $b$ \citep{FederrathKlessenSchmidt2008,FederrathDuvalKlessenSchmidtMacLow2010}. The three dotted lines show Eq.~(\ref{eq:b}) for three representative driving cases: purely solenoidal driving ($b=1/3$, gold), naturally-mixed driving ($b\sim0.4$, blue), and purely compressive driving ($b=1$, purple); see Eq.~(\ref{eq:bvar}). Numerical simulations are shown as symbols (with the color indicating the applied driving mode: sol, mix or comp): from \citet{FederrathKlessenSchmidt2008,FederrathDuvalKlessenSchmidtMacLow2010} (diamonds), \citet{PriceFederrathBrunt2011} (pentagon), \citet{MolinaEtAl2012} (squares), \citet{KonstandinEtAl2012ApJ} (stars), \citet{NolanFederrathSutherland2015} (triangles), and \citet{FederrathBanerjee2015} (upside-down triangle). The black crosses are measurements in the Milky Way spiral-arm clouds Taurus \citep{Brunt2010}, GRSMC43.30-0.33 \citep{GinsburgFederrathDarling2013}, and IC5146 \citep{PadoanJonesNordlund1997}. Our measurement for {\brick} is shown as the red circle with $1\sigma$ uncertainties drawn from the PDFs in the top and right panels. This indicates solenoidal driving of the turbulence in {\brick}, i.e., $b<0.4$. By contrast, all three spiral-arm clouds show a significant compressive driving component, $b>0.4$.
}
\label{fig:b}
\end{figure*}

Figure~\ref{fig:b} shows a graphical representation of the density dispersion--Mach number relation given by Equation~(\ref{eq:b}). In order to put our result for the CMZ cloud {\brick} in context, we include three spiral-arm clouds in Figure~\ref{fig:b}, for which the density dispersion--Mach number relation was measured in previous works. Measurements in Taurus were obtained in \citet{Brunt2010}, with corrections to the Mach number estimate in \citet{KainulainenTan2013}, and combined here with an Alfv\'en Mach number of $\macha>1$ estimated for the gas inside the cloud \citep{HeyerBrunt2012}.\footnote{As a reasonable estimate of $\macha$ in Taurus, we adopt $\macha=2$ and plot in Fig.~\ref{fig:b} the horizontal error bars from the lower limit ($\macha=1$) to 4 times this value ($\macha=4$). Given the low Alfv\'en Mach number in the periphery of Taurus \citep[$\macha\sim0.5$; see][]{HeyerBrunt2012}, it is unlikely that $\macha$ could reach values higher than $\macha=4$ inside Taurus. We further include the uncertainty of the sonic Mach number into the horizontal error bars.} The Galactic Ring Survey molecular cloud GRSMC43.30-0.33 data are from \citet{GinsburgFederrathDarling2013} and the IC5146  data are from \citet{PadoanJonesNordlund1997}. For GRSMC43.30-0.33 and IC5146, we had to neglect the magnetic field (assumed $\macha\to\infty$), because there are no measurements of $B_\mathrm{turb}$ available for these clouds. However, we emphasize that including a realistic magnetic field strength $B_\mathrm{turb}>0$ will always lead to higher values of the driving parameter $b$. The data points for GRSMC43.30-0.33 and IC5146 shown in Figure~\ref{fig:b} therefore represent lower limits of $b$.

We see that all three available spiral-arm clouds have a significant compressive driving component, i.e., they have $b$ parameters exceeding the natural driving mixture, $b>0.4$, given by the blue dotted line \citep{FederrathDuvalKlessenSchmidtMacLow2010}.
However, {\brick} (shown as the red circle in Fig.~\ref{fig:b}) has a significantly lower density dispersion $\sigrho$ and thus significantly lower $b$ than any of the three solar neighborhood clouds. Our measurement of $b=0.22\pm0.12$ indicates solenoidal driving of the turbulence ($b<0.4$). Given the inherent observational uncertainties we rule out mixed driving in favor of solenoidal driving at $1\sigma$ confidence level. We speculate that the most likely physical driver causing this solenoidal driving mode in {\brick} are shearing motions in the CMZ. This is consistent with the large-scale velocity gradient across {\brick} that we saw in Figure~\ref{fig:velocities}, which Kruijssen et al. (in preparation) show rigorously is caused by the shear that {\brick} experienced during its recent pericenter passage \citep{LongmoreEtAl2013b,KruijssenDaleLongmore2015}.

If {\brick} is representative of clouds in the CMZ, then we expect/predict that turbulence is generally driven solenoidally by shear in the CMZ and possibly in the central parts of other galaxies as well \citep{KruijssenLongmore2013,MartigEtAl2013,DavisEtAl2014}. The dominant driver of the turbulence in such environments would be shear \citep[as assumed in][]{KrumholzKruijssen2015}. With our direct measurement of the driving parameter $b=0.22\pm0.12$ in the CMZ, we provide an independent confirmation that shear is a strong turbulence driver in {\brick}.\footnote{We note that the shear can only maintain the turbulence as long as the rotation curve allows it. At a galacto-centric radius of  about $100\,\pc$ (i.e., where {\brick} currently resides), the rotation curve reaches a shear minimum. In the \citet{KrumholzKruijssen2015,KrumholzKruijssenCrocker2016} model, this is why the star formation is episodic: irrespective of the turbulence driving, eventually the gas will hit the shear minimum and collapse to form stars and drive feedback.} This solenoidal driving mode might cause (or at least contribute to) the low SFRs observed in the CMZ \citep{LongmoreEtAl2013a} compared to spiral-am clouds, where the turbulence driving is significantly more compressive (cf.~Fig.~\ref{fig:b}). Indeed, simulations and theoretical models of the SFR show that solenoidal driving can reduce the SFR by an order of magnitude compared to compressive driving \citep{FederrathKlessen2012,PadoanEtAl2014}. More measurements of $b$ are needed in different environments to understand which turbulence drivers are dominant in different physical conditions (e.g., spiral-arm clouds vs.~CMZ, low redshift vs.~high redshift, etc.).

\section{Discussion} \label{sec:caveats}

\subsection{Suppression of dense cores in {\brick}}

Interferometric molecular line and dust emission observations by \citet{KauffmannPillaiZhang2013} and \citet{RathborneEtAl2014} showed a lack of dense cores of significant mass and density in {\brick}, thus providing a possible explanation for the low star formation activity in the CMZ cloud. However, this does not explain what causes the lack of dense cores. Here we find a possible reason, namely that the turbulence is driven solenoidally in {\brick} by large-scale shear, which can suppress the formation of dense cores and reduces the SFR.

\subsection{The star formation rate in {\brick}}

\citet{KruijssenEtAl2014} and \citet{RathborneEtAl2014} showed that the volume density threshold for star formation is several orders of magnitude higher in the CMZ compared to clouds in the solar neighborhood. We now compute the critical density and the SFR for {\brick}. Based on the values derived in Table~\ref{tab:brick} and adopting the \citet{KrumholzMcKee2005} or \citet{PadoanNordlund2011} model for the critical density with the best-fit theory parameter $\phi_x=0.17\pm0.02$ \citep[the ratio of sonic to Jeans scale at the critical density; see Eq.~(10) and Tab.~3 in][]{FederrathKlessen2012}, we find the log-normalized critical density threshold \citep[Eq.~20 in][]{FederrathKlessen2012},
\begin{equation} \label{eq:scrit}
s_\mathrm{crit} = \ln\left[\frac{\pi^2}{5}\phi_x^2 \alpha_\mathrm{vir,tot} \mach^2 \frac{1}{1+\beta^{-1}}\right] = 2.3\pm1.2.
\end{equation}
Note that in this Equation for the critical density of star formation, we used the total (turbulence+shear) virial parameter instead of the purely turbulent one, because shear contributes to reducing the star formation potential of the cloud.

Equation~(\ref{eq:scrit}) leads to a critical volume density threshold of $n_\mathrm{crit} = n_0\exp(s_\mathrm{crit}) = 1.0\times10^5\,\cm^{-3}$, about 1--2 orders of magnitude higher than in nearby clouds. However, this alone does not explain the low SFR in {\brick}, because the gas densities are equally elevated by 1--2 orders of magnitude. Relevant for the predicted SFR based on models of supersonic MHD turbulence is not $n_\mathrm{crit}$, but the log-normalized critical density ($s_\mathrm{crit}$) given by Equation~(\ref{eq:scrit}), which does not depend on the average density $n_0$ \citep{FederrathKlessen2012,PadoanEtAl2014}. Indeed, the theory is fully determined by four dimensionless physical parameters of the cloud, namely the virial parameter, the sonic Mach number, the turbulence driving parameter, and the plasma beta \citep{FederrathKlessen2012}. These four parameters define the multi-freefall model \citep{HennebelleChabrier2011} for the dimensionless SFR per freefall time given by \citep[Eq.~41 in][]{FederrathKlessen2012},
\begin{equation} \label{eq:sfrff}
\sfrff = \frac{\eps}{2\phi_t} \exp\left(\frac{3}{8}\sigma_s^2\right) \left[1+\mathrm{erf}\left(\frac{\sigma_s^2-\s_\mathrm{crit}}{\sqrt{2\sigma_s^2}}\right)\right]
\end{equation}
with the log-normalized density variance \citep[Eq.~4 in][]{FederrathKlessen2012},
\begin{equation} \label{eq:sigs}
\sigma_s^2 = \ln\left[1+b^2\mach^2\beta/(\beta+1)\right].
\end{equation}
Using our derived parameters $\alpha_\mathrm{vir,tot}$, $\mach$, $b$, and $\beta$ for {\brick} from Table~\ref{tab:brick}, and combined with the best-fit theory parameter $1/\phi_t=0.46\pm0.06$ \citep[from Tab.~3 in][]{FederrathKlessen2012} and the core-to-star efficiency $\eps=0.5$ \citep{FederrathEtAl2014}, we find an SFR per freefall time of $\sfrff=0.042\pm0.030$ or an absolute SFR,
\begin{equation} \label{eq:sfr}
\mathrm{SFR}=\sfrff  \, M / t_\mathrm{ff} = (1.1\pm0.8)\times10^{-2}\,\msol\,\yr^{-1}.
\end{equation}

The key point is that the same theoretical model predicts $\mathrm{SFR}=7.6\times10^{-2}\,\msol\,\yr^{-1}$ if a turbulence driving parameter $b=0.5$ is used, which is typical for clouds in the solar neighborhood (see Fig.~\ref{fig:b}). We see that this is a factor of $6.9$ higher than what we derived in Equation~(\ref{eq:sfr}) based on our measured $b=0.22$. This demonstrates that the driving of the turbulence is a critical parameter for the SFR of {\brick}.

\subsection{Turbulent versus ordered magnetic field} \label{sec:turb_vs_ordered_B}

We emphasize that the \emph{turbulent} plasma $\beta$ (not the total plasma $\beta$), enters the theoretical models for the sonic scale, turbulence driving parameter, critical density for star formation, and turbulent density dispersion, given by Equations~(\ref{eq:ls}), (\ref{eq:b}), (\ref{eq:scrit}), and~(\ref{eq:sigs}), respectively. As explained in \citet{FederrathKlessen2012}, these equations are not valid if one inserts the total (turbulent+ordered) plasma $\beta$ in the presence of a strong ordered magnetic field component. This is because the equations were derived by adding the turbulent pressure to the thermal pressure, thus only considering the effect of the turbulent magnetic field. This is why we derived the turbulent magnetic field component of {\brick} in Section~\ref{sec:mag}, which yields the turbulent plasma $\beta$ entering Equations~(\ref{eq:ls}), (\ref{eq:b}), (\ref{eq:scrit}), and~(\ref{eq:sigs}).

\subsection{Comparison with simulations of {\brick}}

\citet{BertramEtAl2015} performed numerical simulations of star formation with the goal to understand the low star formation efficiency in {\brick}. They primarily varied the virial parameter of their model clouds from 0.5 to 8. We measured the total (turbulence+shear) virial parameter in {\brick} and find $\alpha_\mathrm{vir,tot}=4.3\pm2.3$. However, \citet{BertramEtAl2015} find that even such high virial parameters still yield too high star formation efficiencies. A possible reason for the persistent high SFR in their simulations is that \citet{BertramEtAl2015} did not use solenoidal driving of the turbulence, which can reduce the SFR by factors of a few as we have shown in the previous subsection \citep{FederrathKlessen2012}.

\subsection{Absorption filaments}

\citet{BallyEtAl2014} found filaments observed in absorption of the HCO$^{+}$ $J=1-0$ line toward {\brick}. Radiative transfer calculations aimed at reproducing the observations show that the absorption filaments are located in gas of less than $10^3\,\cm^{-3}$ \citep{BallyEtAl2014}. This is low density compared to {\brick}, where the gas densities are $\sim10^4\,\cm^{-3}$ \citep[][Tab.~\ref{tab:brick}]{RathborneEtAl2014b,RathborneEtAl2014}. Thus, \citet{BallyEtAl2014} concluded that the absorption filaments seen in their study may be located close to the surface of {\brick} or in front of {\brick}. Here instead we study filaments identified in the ALMA 3~mm dust continuum distribution, primarily tracing material \emph{inside} {\brick}. \citet{BallyEtAl2014} estimated the H$_2$ column densities of their absorption filaments to be only $6\times10^{20}\,\cm^{-2}$, more than two orders of magnitude lower than the column densities we find here for the dust-continuum filaments (cf.~Fig.~\ref{fig:filprof}).

\subsection{Caveats and limitations}

\subsubsection{Uncertainties in the column density}

The column density maps shown in Figure~\ref{fig:coldensimage} were produced in \citet{LongmoreEtAl2012} and \citet{RathborneEtAl2014} (see Sec.~\ref{sec:observations}). The pure \emph{Herschel} map was derived by modeling the spectral energy distribution (SED) using data from 5 \emph{Herschel} wavelengths obtained with Hi-GAL. To recover the large-scale emission in the ALMA interferometer data, the $500\,\mu\mathrm{m}$ dust continuum emission from \emph{Herschel} was scaled to what is expected at the ALMA 3~mm continuum emission, assuming a graybody where the flux scales as $\nu^{\beta_\mathrm{SED}+2}$ with a global spectral index $\beta_\mathrm{SED}=1.2\pm0.1$ and a global dust temperature $T_\mathrm{dust}=20\pm1\,\mathrm{K}$. \citet{RathborneEtAl2014} estimated the uncertainties following this procedure to be of the order of 10\% in the column density, if the dust temperature and spectral index are fixed and only statistical uncertainties are taken into account. However, the systematic uncertainties in scaling the flux from \emph{Herschel} to the ALMA 3~mm continuum emission actually introduced uncertainties by a factor of 2 in the average column density $N_0$. We obtained this factor 2 uncertainty by comparing $N_0$ in the \emph{Herschel} column density map from \citet{LongmoreEtAl2012} with the $N_0$ in the combined ALMA+\emph{Herschel} map from \citet{RathborneEtAl2014}, shown in Figure~\ref{fig:coldensimage}. Since the \emph{Herschel} map was obtained by SED modeling with data from 5 wavelengths, it provides a well-calibrated column density map. Thus, in order to derive global properties such as the average column density and mass of {\brick} we use the pure \emph{Herschel} map. For extracting filamentary structures, we use the high-resolution ALMA+\emph{Herschel} map \citep{RathborneEtAl2014}. The consequence of the uncertainty in $N_0$ is that the absolute calibration of the column density profiles shown in Figure~\ref{fig:filprof} is also uncertain by a factor of 2, but the derived filament width is independent of $N_0$ (because $N_0$ merely shifts the filament profiles up or down in $N$, but leaves the width unchanged). The normalized column density PDF in \citet{RathborneEtAl2014} and the derived $\sigma_{\eta}$, $\sigma_{N/N_0}$ and $\sigma_{\rho/\rho_0}$ in Table~\ref{tab:brick} are also not affected by the uncertainty in $N_0$. This is because these quantities are defined such that $N_0$ is divided out, so they merely quantify the column- and volume-density contrast, independent of $N_0$.

\subsubsection{Correlation of dust and molecular line emission}

\citet{RathborneEtAl2015} investigated the correlation between the dust emission and 17 molecular line tracers observed toward {\brick}. For most of the molecular tracers, they find a lack of correlation. The best overall correlation is provided by the HNCO line, which is why we use it here to measure the global velocity gradient and velocity dispersion (cf.~Fig.~\ref{fig:velocities}). While the HNCO line provides good coverage and is sufficiently optically thin to trace the global kinematics of {\brick} well, the \emph{local} correlations are often rather poor. This caveat prevents us from studying the detailed velocity structure along the LOS toward each individual filament identified in Figure~\ref{fig:coldensimage}. Previous studies of filamentary structures emphasize the importance of correlation between the column density and velocity structure \citep{HacarEtAl2013,MoeckelBurkert2015,KainulainenEtAl2016,HacarEtAl2016,Federrath2016,SmithEtAl2016}. This must ultimately be done with more reliable line tracers than currently available.

\subsubsection{Filaments in 2D versus filaments in 3D} \label{sec:fil_proj_effects}

Filaments in a 2D projected image of a cloud do not necessarily correspond to filaments in the 3D position-position-position (PPP) space \citep[e.g.,][]{SmithGloverKlessen2014,FernandezLopez2014,LeeEtAl2014}. Thus, the filaments identified in Figure~\ref{fig:coldensimage} only correspond to filaments in projection, while they may actually consist of sub-filaments extending along the LOS. In order to separate contributions from multiple filaments along the LOS in position-position-velocity (PPV) space, one would need to correlate the column density structure with kinematic information from molecular line tracers. However, we currently do not have sufficiently good line tracers to test the correlation in PPV space (see previous subsection). Note that even if we had access to reliable information in PPV space, we would still not be able to separate filaments in PPP space. However, here we are primarily interested in the \emph{statistical averages over all the filaments}, in particular their average width and column density (see Fig.~\ref{fig:filprof}). \citet{SmithGloverKlessen2014} compared the filament widths obtained in 2D versus 3D and find that the mean 3D filament width is (on average) a factor of 2 smaller than the 2D filament width. This is because multiple filaments along the LOS can blend together in the 2D projection. Using a relatively large fit range can therefore overestimate the intrinsic filament width. \citet{Federrath2016} therefore recommended to use a fitting technique that is most sensitive to the peak of the filament profile and reduces the weight of contributions from the wings of the profile (where the LOS blending of other filaments can broaden the profile). If sufficient angular resolution is available this fitting procedure can minimize the effect of the broadening. Nevertheless, we caution that individual filaments identified in Figure~\ref{fig:coldensimage} do not necessarily correspond to coherent and individual structures in the 3D space of {\brick}. Based on the simulations in \citet{SmithGloverKlessen2014} and their comparison of filament detection in 2D and 3D, we apply a factor of 2 uncertainty to our measured filament width.

\citet{MarshEtAl2016} recently identified an elongated structure in the column density map of {\brick} based on \emph{Herschel} data. Since the resolution of \emph{Herschel} is not sufficient to resolve the structure of filaments down to $\lesssim0.1\,\pc$ (see Fig.~\ref{fig:coldensimage}), it is possible that the elongated structure identified in \citet{MarshEtAl2016} actually consists of multiple sub-structures.

\subsubsection{Thermal structure of {\brick}}

The theoretical models for the sonic scale (filament widths) and the standard deviation of density fluctuations, Equations~(\ref{eq:ls}) and~(\ref{eq:b}) respectively, both rest on the assumption of isothermal turbulence, i.e., gas at constant temperature. {\brick} has gas temperature variations ranging from as low as $50\,\mathrm{K}$ up to $340\,\mathrm{K}$ \citep{LisEtAl2001,AoEtAl2013,MillsMorris2013,BallyEtAl2014,GinsburgEtAl2016}. While this is the total range of gas temperature variations across {\brick}, we here only need an estimate of the \emph{average} global gas temperature of {\brick}. We use an average gas temperature of $T=100\pm50\,\mathrm{K}$, based on measurements from the literature (see Tab.~\ref{tab:brick}). However, we emphasize that our results are not sensitive to the exact choice of gas temperature, because the sound speed $\cs\propto T^{1/2}$ entering Equations~(\ref{eq:ls}) and~(\ref{eq:b}) nearly cancels out. This is because both $\mach=\sigma_v/\cs$ and $\beta=2\cs^2/\va^2$ depend on $\cs$. In order to see that, consider small values of $\beta$ as applicable to {\brick} (see Tab.~\ref{tab:brick}) and expand the factor $(1+\beta^{-1})\approx\beta^{-1}$ in Equations~(\ref{eq:ls}) and~(\ref{eq:b}). We see that in the limit $\beta\to0$, the sound speed exactly cancels in both equations. Here we have small $\beta$ instead of $\beta\to0$, so $\cs$ does not cancel out exactly, but almost, such that the end results for the sonic scale $\ls$ and the driving parameter $b$ do not sensitively depend on the choice of sound speed and thus they do not significantly depend on the gas temperature of {\brick}.

\section{Summary and conclusions} \label{sec:conclusions}

We measure and derive new physical parameters for the CMZ cloud {\brick}, which give insight into the filament properties and the turbulence driving mode dominating the cloud and possibly galaxy-center clouds in general. Our measurements and results are summarized in Table~\ref{tab:brick}. Here we list the most important results and conclusions:

\begin{enumerate}

\item Using the DisPerSE filament detection algorithm, we find 11 high-S/N filaments in the dense gas of {\brick} (see Fig.~\ref{fig:coldensimage}). Located along some of these filaments are three over-dense regions with a column density exceeding $2.5\times10^{23}\,\cm^{-2}$. As shown in previous studies, one of these cores has a water maser, which may indicate local active star formation. We find that the filling fraction of these cores is only 0.1\% of the total area of {\brick}, indicating inefficient dense-core and star formation.

\item We construct the average radial profile of the filaments and find a typical filament column density of $\sim10^{23}\,\cm^{-2}$, which is an order of magnitude higher than the average filament column density observed in nearby spiral-arm clouds. We measure an average width of $W_\mathrm{fil}=0.17\pm0.08\,\pc$ (see Fig.~\ref{fig:filprof}).

\item We find that the filament width does not significantly depend on the orientation of the filaments with respect to the ordered magnetic field component in \brick.

\item Based on the column density PDF analyzed in \citet{RathborneEtAl2014} and combined with the column density power spectrum, we reconstruct the volume density dispersion, $\sigrho=1.3\pm0.5$, using the method developed in \citet{BruntFederrathPrice2010a}.

\item Analyzing the spatial distribution of the HNCO intensity-weighted velocity, we see a strong large-scale velocity gradient across the whole cloud, which is likely associated with strong shearing motions (Kruijssen et al., in preparation). We subtract the large-scale gradient in order to obtain the distribution of turbulent velocities. From the Gaussian shape of the velocity PDF (Fig.~\ref{fig:velocities}), we find a turbulent velocity dispersion of $\sigma_{v,\mathrm{1D}}=3.9\pm0.1\,\km\,\s^{-1}$, which is significantly smaller than the total velocity dispersion ($8.8\pm0.2\,\km\,\s^{-1}$).

\item Using magnetohydrodynamical turbulence simulations that take the measured turbulent velocity dispersion and the total (ordered+turbulent) magnetic field strength $B_\mathrm{tot}=2.2\,(0.9)\,\mathrm{m}\Gauss$ adapted from \citet{PillaiEtAl2015} as input, we determine the \emph{turbulent} magnetic field component $B_\mathrm{turb}=130\pm50\,\mu\Gauss$ (Fig.~\ref{fig:magnetic}). Given the velocity dispersion and strong ordered field in {\brick}, our simulations show that $B_\mathrm{turb}$ can only grow to $\lesssim\,B_\mathrm{tot}/10$.

\item Using $B_\mathrm{turb}$ and adding the gas temperature $T=100\pm50\,\mathrm{K}$ constrained in the literature, we derive the sound speed, the Alfv{\'e}n speed and the ratio of thermal to magnetic pressure, plasma $\beta$ (Tab.~\ref{tab:brick}). Using these measurements, we derive a 3D turbulent sonic Mach number of $\mach=11\pm3$ and a turbulent Alfv{\'e}n Mach number of $\macha=4.6\pm2.1$ for {\brick}.

\item We measure the effective cloud diameter $L=4.7\pm0.1\,\pc$ and combine it with the Mach number and plasma $\beta$ to derive the sonic scale $\ls$ of the turbulence in {\brick}. We find $\ls=L\mach^{-2}(1+\beta^{-1})=0.15\pm0.11\,\pc$, in agreement with our measurement of the filament width, $W_\mathrm{fil}=0.17\pm0.08\,\pc$. This supports the idea that the filament width is determined by the sonic scale, Equation~(\ref{eq:ls}), both in the CMZ and in spiral-arm clouds \citep{Federrath2016}. We caution that Equation~(\ref{eq:ls}) strictly only applies to the filament populations perpendicular to the ordered magnetic field; however, we find similar widths for parallel and perpendicular filaments (see Fig.~\ref{fig:filprof}).

\item Our results imply that the filament width in {\brick} is similar to the filament width in nearby clouds, despite the orders-of-magnitude difference in some physical parameters of nearby clouds compared to the CMZ. The reason behind the similarity in $W_\mathrm{fil}$ is the sonic scale, Equation~(\ref{eq:ls}). It depends only on $L$, $\mach=\sigma_{v,\mathrm{3D}}/\cs$ and $\beta=p_\mathrm{thermal}/p_\mathrm{magnetic}$. While the thermal and magnetic pressure are both an order of magnitude higher in {\brick} compared to clouds in solar neighborhood, the ratio (plasma $\beta\sim0.3$) is similar in both environments. The same applies for the sonic Mach number---both $\sigma_{v,\mathrm{3D}}$ and $\cs$ are individually enhanced in {\brick} by factors of a few, but their ratio ($\mach\sim10$) is again similar to nearby clouds \citep{SchneiderEtAl2013}.

\item Using the reconstructed volume density dispersion $\sigrho$ together with $\mach$ and $\beta$ allows us to derive the driving mode parameter $b$ of the turbulence, following Equations~(\ref{eq:b}) and~(\ref{eq:bvar}). We find $b=\sigrho\mach^{-1}(1+\beta^{-1})^{1/2}=0.22\pm0.12$, indicating solenoidal driving in {\brick}.

\item We argue that the solenoidal driving in this Galactic-Center cloud is caused by strong shear, in agreement with the strong large-scale velocity gradient (c.f.~Fig.~\ref{fig:velocities}) and with detailed numerical simulations of CMZ clouds. We speculate that this solenoidal mode of turbulence driving might be the typical driving mode in the centers of galaxies, because of the enhanced shear in such environments. The solenoidal (shearing) mode of turbulence might explain the low SFRs observed in the CMZ compared to spiral-arm clouds, where the driving appears to have a significantly more compressive component, $b>0.4$ (see Fig.~\ref{fig:b}). Using SFR theory based on MHD turbulence, we find that $b=0.22$ yields a factor of $6.9$ lower SFR compared to $b=0.5$, emphasizing the role of the turbulence driving parameter.

\end{enumerate}

\acknowledgements
We thank A.~Ginsburg for discussions on the thermal structure and T.~Pillai and J.~Kauffmann for discussions on the ordered magnetic field in {\brick}. We further thank T.~Csengeri, R.~Klessen, V.~Ossenkopf and N.~Schneider for discussions on noise, foreground corrections and uncertainties in column-density PDFs constructed from interferometric data. We thank the anonymous referees for their prompt and constructive reports.
C.F.~acknowledges funding provided by the Australian Research Council's Discovery Projects (grant~DP150104329). C.F.~further acknowledges supercomputing time provided by the J\"ulich Supercomputing Centre (grant hhd20), the Leibniz Rechenzentrum and the Gauss Centre for Supercomputing (grants~pr32lo, pr48pi and GCS Large-scale project~10391), the Partnership for Advanced Computing in Europe (PRACE grant pr89mu), the Australian National Computational Infrastructure (grant~ek9), and the Pawsey Supercomputing Centre with funding from the Australian Government and the Government of Western Australia.
J.M.D.K.~gratefully acknowledges financial support in the form of a Gliese Fellowship and an Emmy Noether Research Group from the Deutsche Forschungsgemeinschaft (DFG), grant number KR4801/1-1.
R.M.C.~is the recipient of an Australian Research Council Future Fellowship (FT110100108).
GG acknowledges support from CONICYT Project~PFB-06.
This work makes use of the following ALMA data: ADS/JAO.ALMA\#2011.0.00217.S. ALMA is a partnership of ESO (representing its member states), NSF (USA), and NINS (Japan), together with NRC (Canada) and NSC and ASIAA (Taiwan), in cooperation with the Republic of Chile. The Joint ALMA Observatory is operated by ESO, AUI/NRAO, and NAOJ.
The simulation software FLASH used in this work was in part developed by the DOE-supported Flash Center for Computational Science at the University of Chicago.

\emph{Facilities:} ALMA, \emph{Herschel}, Mopra.


\begin{thebibliography}{195}
\expandafter\ifx\csname natexlab\endcsname\relax\def\natexlab#1{#1}\fi

\bibitem[{{Andr{\'e}} {et~al.}(2014){Andr{\'e}}, {Di Francesco},
  {Ward-Thompson}, {Inutsuka}, {Pudritz}, \& {Pineda}}]{AndreEtAl2014}
{Andr{\'e}}, P., {Di Francesco}, J., {Ward-Thompson}, D., {et~al.} 2014,
  Protostars and Planets VI, 27

\bibitem[{{Andr{\'e}} {et~al.}(2010){Andr{\'e}}, {Men'shchikov}, {Bontemps},
  {K{\"o}nyves}, {Motte}, {Schneider}, {Didelon}, {Minier}, {Saraceno},
  {Ward-Thompson}, {di Francesco}, {White}, {Molinari}, {Testi}, {Abergel},
  {Griffin}, {Henning}, {Royer}, {Mer{\'{\i}}n}, {Vavrek}, {Attard},
  {Arzoumanian}, {Wilson}, {Ade}, {Aussel}, {Baluteau}, {Benedettini},
  {Bernard}, {Blommaert}, {Cambr{\'e}sy}, {Cox}, {di Giorgio}, {Hargrave},
  {Hennemann}, {Huang}, {Kirk}, {Krause}, {Launhardt}, {Leeks}, {Le Pennec},
  {Li}, {Martin}, {Maury}, {Olofsson}, {Omont}, {Peretto}, {Pezzuto}, {Prusti},
  {Roussel}, {Russeil}, {Sauvage}, {Sibthorpe}, {Sicilia-Aguilar}, {Spinoglio},
  {Waelkens}, {Woodcraft}, \& {Zavagno}}]{AndreEtAl2010}
{Andr{\'e}}, P., {Men'shchikov}, A., {Bontemps}, S., {et~al.} 2010, \aap, 518,
  L102

\bibitem[{{Ao} {et~al.}(2013){Ao}, {Henkel}, {Menten}, {Requena-Torres},
  {Stanke}, {Mauersberger}, {Aalto}, {M{\"u}hle}, \& {Mangum}}]{AoEtAl2013}
{Ao}, Y., {Henkel}, C., {Menten}, K.~M., {et~al.} 2013, \aap, 550, A135

\bibitem[{{Arce} {et~al.}(2011){Arce}, {Borkin}, {Goodman}, {Pineda}, \&
  {Beaumont}}]{ArceEtAl2011}
{Arce}, H.~G., {Borkin}, M.~A., {Goodman}, A.~A., {Pineda}, J.~E., \&
  {Beaumont}, C.~N. 2011, \apj, 742, 105

\bibitem[{{Arzoumanian} {et~al.}(2011){Arzoumanian}, {Andr{\'e}}, {Didelon},
  {K{\"o}nyves}, {Schneider}, {Men'shchikov}, {Sousbie}, {Zavagno}, {Bontemps},
  {di Francesco}, {Griffin}, {Hennemann}, {Hill}, {Kirk}, {Martin}, {Minier},
  {Molinari}, {Motte}, {Peretto}, {Pezzuto}, {Spinoglio}, {Ward-Thompson},
  {White}, \& {Wilson}}]{ArzoumanianEtAl2011}
{Arzoumanian}, D., {Andr{\'e}}, P., {Didelon}, P., {et~al.} 2011, \aap, 529, L6

\bibitem[{{Bally} {et~al.}(2014){Bally}, {Rathborne}, {Longmore}, {Jackson},
  {Alves}, {Bressert}, {Contreras}, {Foster}, {Garay}, {Ginsburg}, {Johnston},
  {Kruijssen}, {Testi}, \& {Walsh}}]{BallyEtAl2014}
{Bally}, J., {Rathborne}, J.~M., {Longmore}, S.~N., {et~al.} 2014, \apj, 795,
  28

\bibitem[{{Balsara} {et~al.}(2001){Balsara}, {Ward-Thompson}, \&
  {Crutcher}}]{BalsaraEtAl2001}
{Balsara}, D., {Ward-Thompson}, D., \& {Crutcher}, R.~M. 2001, \mnras, 327, 715

\bibitem[{{Balsara} {et~al.}(2004){Balsara}, {Kim}, {Mac Low}, \&
  {Mathews}}]{BalsaraEtAl2004}
{Balsara}, D.~S., {Kim}, J., {Mac Low}, M., \& {Mathews}, G.~J. 2004, \apj,
  617, 339

\bibitem[{{Banerjee} {et~al.}(2007){Banerjee}, {Klessen}, \&
  {Fendt}}]{BanerjeeKlessenFendt2007}
{Banerjee}, R., {Klessen}, R.~S., \& {Fendt}, C. 2007, \apj, 668, 1028

\bibitem[{{Benedettini} {et~al.}(2015){Benedettini}, {Schisano}, {Pezzuto},
  {Elia}, {Andr{\'e}}, {K{\"o}nyves}, {Schneider}, {Tremblin}, {Arzoumanian},
  {di Giorgio}, {Di Francesco}, {Hill}, {Molinari}, {Motte}, {Nguyen-Luong},
  {Palmeirim}, {Rivera-Ingraham}, {Roy}, {Rygl}, {Spinoglio}, {Ward-Thompson},
  \& {White}}]{BenedettiniEtAl2015}
{Benedettini}, M., {Schisano}, E., {Pezzuto}, S., {et~al.} 2015, \mnras, 453,
  2036

\bibitem[{{Berkhuijsen} \& {Fletcher}(2008)}]{BerkhuijsenFletcher2008}
{Berkhuijsen}, E.~M., \& {Fletcher}, A. 2008, \mnras, 390, L19

\bibitem[{{Berkhuijsen} \& {Fletcher}(2015)}]{BerkhuijsenFletcher2015}
---. 2015, \mnras, 448, 2469

\bibitem[{{Bertram} {et~al.}(2015){Bertram}, {Glover}, {Clark}, \&
  {Klessen}}]{BertramEtAl2015}
{Bertram}, E., {Glover}, S.~C.~O., {Clark}, P.~C., \& {Klessen}, R.~S. 2015,
  \mnras, 451, 3679

\bibitem[{{Breen} \& {Ellingsen}(2011)}]{BreenEllingsen2011}
{Breen}, S.~L., \& {Ellingsen}, S.~P. 2011, \mnras, 416, 178

\bibitem[{{Breitschwerdt} {et~al.}(2009){Breitschwerdt}, {de Avillez}, {Fuchs},
  \& {Dettbarn}}]{BreitschwerdtEtAl2009}
{Breitschwerdt}, D., {de Avillez}, M.~A., {Fuchs}, B., \& {Dettbarn}, C. 2009,
  Space Science Reviews, 143, 263

\bibitem[{{Brunt}(2010)}]{Brunt2010}
{Brunt}, C.~M. 2010, \aap, 513, A67

\bibitem[{{Brunt} {et~al.}(2010{\natexlab{a}}){Brunt}, {Federrath}, \&
  {Price}}]{BruntFederrathPrice2010b}
{Brunt}, C.~M., {Federrath}, C., \& {Price}, D.~J. 2010{\natexlab{a}}, \mnras,
  405, L56

\bibitem[{{Brunt} {et~al.}(2010{\natexlab{b}}){Brunt}, {Federrath}, \&
  {Price}}]{BruntFederrathPrice2010a}
---. 2010{\natexlab{b}}, \mnras, 403, 1507

\bibitem[{{Burkert} \& {Bodenheimer}(2000)}]{BurkertBodenheimer2000}
{Burkert}, A., \& {Bodenheimer}, P. 2000, \apj, 543, 822

\bibitem[{{Burkhart} {et~al.}(2009){Burkhart}, {Falceta-Gon{\c c}alves},
  {Kowal}, \& {Lazarian}}]{BurkhartEtAl2009}
{Burkhart}, B., {Falceta-Gon{\c c}alves}, D., {Kowal}, G., \& {Lazarian}, A.
  2009, \apj, 693, 250

\bibitem[{{Burkhart} \& {Lazarian}(2012)}]{BurkhartLazarian2012}
{Burkhart}, B., \& {Lazarian}, A. 2012, \apjl, 755, L19

\bibitem[{{Carroll} {et~al.}(2010){Carroll}, {Frank}, \&
  {Blackman}}]{CarrollFrankBlackman2010}
{Carroll}, J.~J., {Frank}, A., \& {Blackman}, E.~G. 2010, \apj, 722, 145

\bibitem[{{Chabrier} \& {Hennebelle}(2011)}]{ChabrierHennebelle2011}
{Chabrier}, G., \& {Hennebelle}, P. 2011, \aap, 534, A106

\bibitem[{{Chabrier} {et~al.}(2014){Chabrier}, {Hennebelle}, \&
  {Charlot}}]{ChabrierEtAl2014}
{Chabrier}, G., {Hennebelle}, P., \& {Charlot}, S. 2014, \apj, 796, 75

\bibitem[{{Chandrasekhar} \& {Fermi}(1953)}]{ChandrasekharFermi1953}
{Chandrasekhar}, S., \& {Fermi}, E. 1953, \apj, 118, 113

\bibitem[{{Chuss} {et~al.}(2003){Chuss}, {Davidson}, {Dotson}, {Dowell},
  {Hildebrand}, {Novak}, \& {Vaillancourt}}]{ChussEtAl2003}
{Chuss}, D.~T., {Davidson}, J.~A., {Dotson}, J.~L., {et~al.} 2003, \apj, 599,
  1116

\bibitem[{{Contreras} {et~al.}(2013){Contreras}, {Rathborne}, \&
  {Garay}}]{ContrerasRathborneGaray2013}
{Contreras}, Y., {Rathborne}, J., \& {Garay}, G. 2013, \mnras, 433, 251

\bibitem[{{Crocker} {et~al.}(2011){Crocker}, {Jones}, {Aharonian}, {Law},
  {Melia}, {Oka}, \& {Ott}}]{CrockerEtAl2011}
{Crocker}, R.~M., {Jones}, D.~I., {Aharonian}, F., {et~al.} 2011, \mnras, 413,
  763

\bibitem[{{Crocker} {et~al.}(2010){Crocker}, {Jones}, {Melia}, {Ott}, \&
  {Protheroe}}]{CrockerEtAl2010}
{Crocker}, R.~M., {Jones}, D.~I., {Melia}, F., {Ott}, J., \& {Protheroe}, R.~J.
  2010, \nat, 463, 65

\bibitem[{{Cunningham} {et~al.}(2009){Cunningham}, {Frank}, {Carroll},
  {Blackman}, \& {Quillen}}]{CunninghamEtAl2009}
{Cunningham}, A.~J., {Frank}, A., {Carroll}, J., {Blackman}, E.~G., \&
  {Quillen}, A.~C. 2009, \apj, 692, 816

\bibitem[{{Cunningham} {et~al.}(2011){Cunningham}, {Klein}, {Krumholz}, \&
  {McKee}}]{CunninghamEtAl2011}
{Cunningham}, A.~J., {Klein}, R.~I., {Krumholz}, M.~R., \& {McKee}, C.~F. 2011,
  \apj, 740, 107

\bibitem[{{D'Agostini}(2004)}]{DAgostini2004}
{D'Agostini}, G. 2004, ArXiv physics/0403086

\bibitem[{{Davis} {et~al.}(2014){Davis}, {Young}, {Crocker}, {Bureau}, {Blitz},
  {Alatalo}, {Emsellem}, {Naab}, {Bayet}, {Bois}, {Bournaud}, {Cappellari},
  {Davies}, {de Zeeuw}, {Duc}, {Khochfar}, {Krajnovi{\'c}}, {Kuntschner},
  {McDermid}, {Morganti}, {Oosterloo}, {Sarzi}, {Scott}, {Serra}, \&
  {Weijmans}}]{DavisEtAl2014}
{Davis}, T.~A., {Young}, L.~M., {Crocker}, A.~F., {et~al.} 2014, \mnras, 444,
  3427

\bibitem[{{Del Sordo} \& {Brandenburg}(2011)}]{DelSordoBrandenburg2011}
{Del Sordo}, F., \& {Brandenburg}, A. 2011, \aap, 528, A145

\bibitem[{{Dobbs} \& {Bonnell}(2008)}]{DobbsBonnell2008}
{Dobbs}, C.~L., \& {Bonnell}, I.~A. 2008, \mnras, 385, 1893

\bibitem[{{Dobbs} {et~al.}(2008){Dobbs}, {Glover}, {Clark}, \&
  {Klessen}}]{DobbsEtAl2008}
{Dobbs}, C.~L., {Glover}, S.~C.~O., {Clark}, P.~C., \& {Klessen}, R.~S. 2008,
  \mnras, 389, 1097

\bibitem[{{Donkov} {et~al.}(2012){Donkov}, {Veltchev}, \&
  {Klessen}}]{DonkovVeltchevKlessen2012}
{Donkov}, S., {Veltchev}, T.~V., \& {Klessen}, R.~S. 2012, \mnras, 423, 889

\bibitem[{{Dotson} {et~al.}(2010){Dotson}, {Vaillancourt}, {Kirby}, {Dowell},
  {Hildebrand}, \& {Davidson}}]{DotsonEtAl2010}
{Dotson}, J.~L., {Vaillancourt}, J.~E., {Kirby}, L., {et~al.} 2010, \apjs, 186,
  406

\bibitem[{{Elmegreen}(2008)}]{Elmegreen2008}
{Elmegreen}, B.~G. 2008, \apj, 672, 1006

\bibitem[{{Elmegreen} \& {Burkert}(2010)}]{ElmegreenBurkert2010}
{Elmegreen}, B.~G., \& {Burkert}, A. 2010, \apj, 712, 294

\bibitem[{{Falgarone} {et~al.}(2009){Falgarone}, {Pety}, \&
  {Hily-Blant}}]{FalgaronePetyHilyBlant2009}
{Falgarone}, E., {Pety}, J., \& {Hily-Blant}, P. 2009, \aap, 507, 355

\bibitem[{{Falgarone} \& {Phillips}(1990)}]{FalgaronePhillips1990}
{Falgarone}, E., \& {Phillips}, T.~G. 1990, \apj, 359, 344

\bibitem[{{Federrath}(2013)}]{Federrath2013}
{Federrath}, C. 2013, \mnras, 436, 1245

\bibitem[{{Federrath}(2015)}]{Federrath2015}
---. 2015, \mnras, 450, 4035

\bibitem[{{Federrath}(2016)}]{Federrath2016}
---. 2016, \mnras, 457, 375

\bibitem[{{Federrath} \& {Banerjee}(2015)}]{FederrathBanerjee2015}
{Federrath}, C., \& {Banerjee}, S. 2015, \mnras, 448, 3297

\bibitem[{{Federrath} {et~al.}(2011{\natexlab{a}}){Federrath}, {Chabrier},
  {Schober}, {Banerjee}, {Klessen}, \& {Schleicher}}]{FederrathEtAl2011PRL}
{Federrath}, C., {Chabrier}, G., {Schober}, J., {et~al.} 2011{\natexlab{a}},
  PhRvL, 107, 114504

\bibitem[{{Federrath} {et~al.}(2008{\natexlab{a}}){Federrath}, {Glover},
  {Klessen}, \& {Schmidt}}]{FederrathGloverKlessenSchmidt2008}
{Federrath}, C., {Glover}, S.~C.~O., {Klessen}, R.~S., \& {Schmidt}, W.
  2008{\natexlab{a}}, PhST, 132, 014025

\bibitem[{{Federrath} \& {Klessen}(2012)}]{FederrathKlessen2012}
{Federrath}, C., \& {Klessen}, R.~S. 2012, \apj, 761, 156

\bibitem[{{Federrath} \& {Klessen}(2013)}]{FederrathKlessen2013}
---. 2013, \apj, 763, 51

\bibitem[{{Federrath} {et~al.}(2008{\natexlab{b}}){Federrath}, {Klessen}, \&
  {Schmidt}}]{FederrathKlessenSchmidt2008}
{Federrath}, C., {Klessen}, R.~S., \& {Schmidt}, W. 2008{\natexlab{b}}, \apjl,
  688, L79

\bibitem[{{Federrath} {et~al.}(2009){Federrath}, {Klessen}, \&
  {Schmidt}}]{FederrathKlessenSchmidt2009}
---. 2009, \apj, 692, 364

\bibitem[{{Federrath} {et~al.}(2010){Federrath}, {Roman-Duval}, {Klessen},
  {Schmidt}, \& {Mac Low}}]{FederrathDuvalKlessenSchmidtMacLow2010}
{Federrath}, C., {Roman-Duval}, J., {Klessen}, R.~S., {Schmidt}, W., \& {Mac
  Low}, M. 2010, \aap, 512, A81

\bibitem[{{Federrath} {et~al.}(2014){Federrath}, {Schr{\"o}n}, {Banerjee}, \&
  {Klessen}}]{FederrathEtAl2014}
{Federrath}, C., {Schr{\"o}n}, M., {Banerjee}, R., \& {Klessen}, R.~S. 2014,
  \apj, 790, 128

\bibitem[{{Federrath} {et~al.}(2011{\natexlab{b}}){Federrath}, {Sur},
  {Schleicher}, {Banerjee}, \&
  {Klessen}}]{FederrathSurSchleicherBanerjeeKlessen2011}
{Federrath}, C., {Sur}, S., {Schleicher}, D.~R.~G., {Banerjee}, R., \&
  {Klessen}, R.~S. 2011{\natexlab{b}}, \apj, 731, 62

\bibitem[{{Fern{\'a}ndez-L{\'o}pez} {et~al.}(2014){Fern{\'a}ndez-L{\'o}pez},
  {Arce}, {Looney}, {Mundy}, {Storm}, {Teuben}, {Lee}, {Segura-Cox}, {Isella},
  {Tobin}, {Rosolowsky}, {Plunkett}, {Kwon}, {Kauffmann}, {Ostriker}, {Tassis},
  {Shirley}, \& {Pound}}]{FernandezLopez2014}
{Fern{\'a}ndez-L{\'o}pez}, M., {Arce}, H.~G., {Looney}, L., {et~al.} 2014,
  \apjl, 790, L19

\bibitem[{{Ferri{\`e}re}(2010)}]{Ferriere2010}
{Ferri{\`e}re}, K. 2010, Astronomische Nachrichten, 331, 27

\bibitem[{{Gaensler} {et~al.}(2011){Gaensler}, {Haverkorn}, {Burkhart},
  {Newton-McGee}, {Ekers}, {Lazarian}, {McClure-Griffiths}, {Robishaw},
  {Dickey}, \& {Green}}]{GaenslerEtAl2011}
{Gaensler}, B.~M., {Haverkorn}, M., {Burkhart}, B., {et~al.} 2011, \nat, 478,
  214

\bibitem[{{Ginsburg} {et~al.}(2013){Ginsburg}, {Federrath}, \&
  {Darling}}]{GinsburgFederrathDarling2013}
{Ginsburg}, A., {Federrath}, C., \& {Darling}, J. 2013, \apj, 779, 50

\bibitem[{{Ginsburg} {et~al.}(2016){Ginsburg}, {Henkel}, {Ao}, {Riquelme},
  {Kauffmann}, {Pillai}, {Mills}, {Requena-Torres}, {Immer}, {Testi}, {Ott},
  {Bally}, {Battersby}, {Darling}, {Aalto}, {Stanke}, {Kendrew}, {Kruijssen},
  {Longmore}, {Dale}, {Guesten}, \& {Menten}}]{GinsburgEtAl2016}
{Ginsburg}, A., {Henkel}, C., {Ao}, Y., {et~al.} 2016, \aap, 586, A50

\bibitem[{{Girichidis} {et~al.}(2014){Girichidis}, {Konstandin}, {Whitworth},
  \& {Klessen}}]{GirichidisEtAl2014}
{Girichidis}, P., {Konstandin}, L., {Whitworth}, A.~P., \& {Klessen}, R.~S.
  2014, \apj, 781, 91

\bibitem[{{Goldbaum} {et~al.}(2011){Goldbaum}, {Krumholz}, {Matzner}, \&
  {McKee}}]{GoldbaumEtAl2011}
{Goldbaum}, N.~J., {Krumholz}, M.~R., {Matzner}, C.~D., \& {McKee}, C.~F. 2011,
  \apj, 738, 101

\bibitem[{{Hacar} {et~al.}(2016){Hacar}, {Kainulainen}, {Tafalla}, {Beuther},
  \& {Alves}}]{HacarEtAl2016}
{Hacar}, A., {Kainulainen}, J., {Tafalla}, M., {Beuther}, H., \& {Alves}, J.
  2016, \aap, 587, A97

\bibitem[{{Hacar} \& {Tafalla}(2011)}]{HacarTafalla2011}
{Hacar}, A., \& {Tafalla}, M. 2011, \aap, 533, A34

\bibitem[{{Hacar} {et~al.}(2013){Hacar}, {Tafalla}, {Kauffmann}, \&
  {Kov{\'a}cs}}]{HacarEtAl2013}
{Hacar}, A., {Tafalla}, M., {Kauffmann}, J., \& {Kov{\'a}cs}, A. 2013, \aap,
  554, A55

\bibitem[{{Hennebelle}(2013)}]{Hennebelle2013}
{Hennebelle}, P. 2013, \aap, 556, A153

\bibitem[{{Hennebelle} \& {Chabrier}(2008)}]{HennebelleChabrier2008}
{Hennebelle}, P., \& {Chabrier}, G. 2008, \apj, 684, 395

\bibitem[{{Hennebelle} \& {Chabrier}(2009)}]{HennebelleChabrier2009}
---. 2009, \apj, 702, 1428

\bibitem[{{Hennebelle} \& {Chabrier}(2011)}]{HennebelleChabrier2011}
---. 2011, \apjl, 743, L29

\bibitem[{{Hennebelle} \& {Chabrier}(2013)}]{HennebelleChabrier2013}
---. 2013, \apj, 770, 150

\bibitem[{{Henshaw} {et~al.}(2016){Henshaw}, {Longmore}, {Kruijssen}, {Davies},
  {Bally}, {Barnes}, {Battersby}, {Burton}, {Cunningham}, {Dale}, {Ginsburg},
  {Immer}, {Jones}, {Kendrew}, {Mills}, {Molinari}, {Moore}, {Ott}, {Pillai},
  {Rathborne}, {Schilke}, {Schmiedeke}, {Testi}, {Walker}, {Walsh}, \&
  {Zhang}}]{HenshawEtAl2016}
{Henshaw}, J.~D., {Longmore}, S.~N., {Kruijssen}, J.~M.~D., {et~al.} 2016,
  \mnras, 457, 2675

\bibitem[{{Heyer} \& {Brunt}(2012)}]{HeyerBrunt2012}
{Heyer}, M.~H., \& {Brunt}, C.~M. 2012, \mnras, 420, 1562

\bibitem[{{Hily-Blant} {et~al.}(2008){Hily-Blant}, {Falgarone}, \&
  {Pety}}]{HilyBlantFalgaronePety2008}
{Hily-Blant}, P., {Falgarone}, E., \& {Pety}, J. 2008, \aap, 481, 367

\bibitem[{{Hopkins}(2012)}]{Hopkins2012b}
{Hopkins}, P.~F. 2012, \mnras, 423, 2037

\bibitem[{{Hopkins}(2013{\natexlab{a}})}]{Hopkins2013IMF}
---. 2013{\natexlab{a}}, \mnras, 430, 1653

\bibitem[{{Hopkins}(2013{\natexlab{b}})}]{Hopkins2013PDF}
---. 2013{\natexlab{b}}, \mnras, 430, 1880

\bibitem[{{Hughes} {et~al.}(2013){Hughes}, {Meidt}, {Schinnerer}, {Colombo},
  {Pety}, {Leroy}, {Dobbs}, {Garc{\'{\i}}a-Burillo}, {Thompson}, {Dumas},
  {Schuster}, \& {Kramer}}]{HughesEtAl2013}
{Hughes}, A., {Meidt}, S.~E., {Schinnerer}, E., {et~al.} 2013, \apj, 779, 44

\bibitem[{{Johnston} {et~al.}(2014){Johnston}, {Beuther}, {Linz}, {Schmiedeke},
  {Ragan}, \& {Henning}}]{JohnstonEtAl2014}
{Johnston}, K.~G., {Beuther}, H., {Linz}, H., {et~al.} 2014, \aap, 568, A56

\bibitem[{{Juvela} {et~al.}(2012){Juvela}, {Ristorcelli}, {Pagani}, {Doi},
  {Pelkonen}, {Marshall}, {Bernard}, {Falgarone}, {Malinen}, {Marton},
  {McGehee}, {Montier}, {Motte}, {Paladini}, {T{\'o}th}, {Ysard}, {Zahorecz},
  \& {Zavagno}}]{JuvelaEtAl2012a}
{Juvela}, M., {Ristorcelli}, I., {Pagani}, L., {et~al.} 2012, \aap, 541, A12

\bibitem[{{Kainulainen} {et~al.}(2009){Kainulainen}, {Beuther}, {Henning}, \&
  {Plume}}]{KainulainenEtAl2009}
{Kainulainen}, J., {Beuther}, H., {Henning}, T., \& {Plume}, R. 2009, \aap,
  508, L35

\bibitem[{{Kainulainen} {et~al.}(2013){Kainulainen}, {Federrath}, \&
  {Henning}}]{KainulainenFederrathHenning2013}
{Kainulainen}, J., {Federrath}, C., \& {Henning}, T. 2013, \aap, 553, L8

\bibitem[{{Kainulainen} {et~al.}(2014){Kainulainen}, {Federrath}, \&
  {Henning}}]{KainulainenFederrathHenning2014}
---. 2014, Science, 344, 183

\bibitem[{{Kainulainen} {et~al.}(2016){Kainulainen}, {Hacar}, {Alves},
  {Beuther}, {Bouy}, \& {Tafalla}}]{KainulainenEtAl2016}
{Kainulainen}, J., {Hacar}, A., {Alves}, J., {et~al.} 2016, \aap, 586, A27

\bibitem[{{Kainulainen} \& {Tan}(2013)}]{KainulainenTan2013}
{Kainulainen}, J., \& {Tan}, J.~C. 2013, \aap, 549, A53

\bibitem[{{Kauffmann} {et~al.}(2008){Kauffmann}, {Bertoldi}, {Bourke}, {Evans},
  \& {Lee}}]{KauffmannEtAl2008}
{Kauffmann}, J., {Bertoldi}, F., {Bourke}, T.~L., {Evans}, II, N.~J., \& {Lee},
  C.~W. 2008, \aap, 487, 993

\bibitem[{{Kauffmann} {et~al.}(2013){Kauffmann}, {Pillai}, \&
  {Zhang}}]{KauffmannPillaiZhang2013}
{Kauffmann}, J., {Pillai}, T., \& {Zhang}, Q. 2013, \apjl, 765, L35

\bibitem[{{Kirk} {et~al.}(2015){Kirk}, {Klassen}, {Pudritz}, \&
  {Pillsworth}}]{KirkEtAl2015}
{Kirk}, H., {Klassen}, M., {Pudritz}, R., \& {Pillsworth}, S. 2015, \apj, 802,
  75

\bibitem[{{Klessen}(2000)}]{Klessen2000}
{Klessen}, R.~S. 2000, \apj, 535, 869

\bibitem[{{Klessen} \& {Hennebelle}(2010)}]{KlessenHennebelle2010}
{Klessen}, R.~S., \& {Hennebelle}, P. 2010, \aap, 520, A17

\bibitem[{{Konstandin} {et~al.}(2012){Konstandin}, {Girichidis}, {Federrath},
  \& {Klessen}}]{KonstandinEtAl2012ApJ}
{Konstandin}, L., {Girichidis}, P., {Federrath}, C., \& {Klessen}, R.~S. 2012,
  \apj, 761, 149

\bibitem[{{K{\"o}nyves} {et~al.}(2015){K{\"o}nyves}, {Andr{\'e}},
  {Men'shchikov}, {Palmeirim}, {Arzoumanian}, {Schneider}, {Roy}, {Didelon},
  {Maury}, {Shimajiri}, {Di Francesco}, {Bontemps}, {Peretto}, {Benedettini},
  {Bernard}, {Elia}, {Griffin}, {Hill}, {Kirk}, {Ladjelate}, {Marsh}, {Martin},
  {Motte}, {Nguy{\^e}n Luong}, {Pezzuto}, {Roussel}, {Rygl}, {Sadavoy},
  {Schisano}, {Spinoglio}, {Ward-Thompson}, \& {White}}]{KonyvesEtAl2015}
{K{\"o}nyves}, V., {Andr{\'e}}, P., {Men'shchikov}, A., {et~al.} 2015, \aap,
  584, A91

\bibitem[{{Kritsuk} {et~al.}(2007){Kritsuk}, {Norman}, {Padoan}, \&
  {Wagner}}]{KritsukEtAl2007}
{Kritsuk}, A.~G., {Norman}, M.~L., {Padoan}, P., \& {Wagner}, R. 2007, \apj,
  665, 416

\bibitem[{{Kritsuk} {et~al.}(2011){Kritsuk}, {Norman}, \&
  {Wagner}}]{KritsukNormanWagner2011}
{Kritsuk}, A.~G., {Norman}, M.~L., \& {Wagner}, R. 2011, \apjl, 727, L20

\bibitem[{{Kruijssen}(2012)}]{Kruijssen2012}
{Kruijssen}, J.~M.~D. 2012, \mnras, 426, 3008

\bibitem[{{Kruijssen} {et~al.}(2015){Kruijssen}, {Dale}, \&
  {Longmore}}]{KruijssenDaleLongmore2015}
{Kruijssen}, J.~M.~D., {Dale}, J.~E., \& {Longmore}, S.~N. 2015, \mnras, 447,
  1059

\bibitem[{{Kruijssen} \& {Longmore}(2013)}]{KruijssenLongmore2013}
{Kruijssen}, J.~M.~D., \& {Longmore}, S.~N. 2013, \mnras, 435, 2598

\bibitem[{{Kruijssen} {et~al.}(2014){Kruijssen}, {Longmore}, {Elmegreen},
  {Murray}, {Bally}, {Testi}, \& {Kennicutt}}]{KruijssenEtAl2014}
{Kruijssen}, J.~M.~D., {Longmore}, S.~N., {Elmegreen}, B.~G., {et~al.} 2014,
  \mnras, 440, 3370

\bibitem[{{Krumholz} \& {Kruijssen}(2015)}]{KrumholzKruijssen2015}
{Krumholz}, M.~R., \& {Kruijssen}, J.~M.~D. 2015, \mnras, 453, 739

\bibitem[{{Krumholz} {et~al.}(2016){Krumholz}, {Kruijssen}, \&
  {Crocker}}]{KrumholzKruijssenCrocker2016}
{Krumholz}, M.~R., {Kruijssen}, J.~M.~D., \& {Crocker}, R.~M. 2016, \mnras,
  submitted (arXiv:1605.02850)

\bibitem[{{Krumholz} {et~al.}(2006){Krumholz}, {Matzner}, \&
  {McKee}}]{KrumholzMatznerMcKee2006}
{Krumholz}, M.~R., {Matzner}, C.~D., \& {McKee}, C.~F. 2006, \apj, 653, 361

\bibitem[{{Krumholz} \& {McKee}(2005)}]{KrumholzMcKee2005}
{Krumholz}, M.~R., \& {McKee}, C.~F. 2005, \apj, 630, 250

\bibitem[{{Lee} {et~al.}(2015){Lee}, {Chang}, \& {Murray}}]{LeeChangMurray2015}
{Lee}, E.~J., {Chang}, P., \& {Murray}, N. 2015, \apj, 800, 49

\bibitem[{{Lee} {et~al.}(2012){Lee}, {Murray}, \&
  {Rahman}}]{LeeMurrayRahman2012}
{Lee}, E.~J., {Murray}, N., \& {Rahman}, M. 2012, \apj, 752, 146

\bibitem[{{Lee} {et~al.}(2014){Lee}, {Fern{\'a}ndez-L{\'o}pez}, {Storm},
  {Looney}, {Mundy}, {Segura-Cox}, {Teuben}, {Rosolowsky}, {Arce}, {Ostriker},
  {Shirley}, {Kwon}, {Kauffmann}, {Tobin}, {Plunkett}, {Pound}, {Salter},
  {Volgenau}, {Chen}, {Tassis}, {Isella}, {Crutcher}, {Gammie}, \&
  {Testi}}]{LeeEtAl2014}
{Lee}, K.~I., {Fern{\'a}ndez-L{\'o}pez}, M., {Storm}, S., {et~al.} 2014, \apj,
  797, 76

\bibitem[{{Li} {et~al.}(2014){Li}, {Goodman}, {Sridharan}, {Houde}, {Li},
  {Novak}, \& {Tang}}]{LiEtAl2014}
{Li}, H.-B., {Goodman}, A., {Sridharan}, T.~K., {et~al.} 2014, Protostars and
  Planets VI, 101

\bibitem[{{Lis} {et~al.}(1994){Lis}, {Menten}, {Serabyn}, \&
  {Zylka}}]{LisEtAl1994}
{Lis}, D.~C., {Menten}, K.~M., {Serabyn}, E., \& {Zylka}, R. 1994, \apjl, 423,
  L39

\bibitem[{{Lis} {et~al.}(2001){Lis}, {Serabyn}, {Zylka}, \& {Li}}]{LisEtAl2001}
{Lis}, D.~C., {Serabyn}, E., {Zylka}, R., \& {Li}, Y. 2001, \apj, 550, 761

\bibitem[{{Lombardi} {et~al.}(2011){Lombardi}, {Alves}, \&
  {Lada}}]{LombardiAlvesLada2011}
{Lombardi}, M., {Alves}, J., \& {Lada}, C.~J. 2011, \aap, 535, A16

\bibitem[{{Longmore} {et~al.}(2012){Longmore}, {Rathborne}, {Bastian}, {Alves},
  {Ascenso}, {Bally}, {Testi}, {Longmore}, {Battersby}, {Bressert}, {Purcell},
  {Walsh}, {Jackson}, {Foster}, {Molinari}, {Meingast}, {Amorim}, {Lima},
  {Marques}, {Moitinho}, {Pinhao}, {Rebordao}, \& {Santos}}]{LongmoreEtAl2012}
{Longmore}, S.~N., {Rathborne}, J., {Bastian}, N., {et~al.} 2012, \apj, 746,
  117

\bibitem[{{Longmore} {et~al.}(2013{\natexlab{a}}){Longmore}, {Kruijssen},
  {Bally}, {Ott}, {Testi}, {Rathborne}, {Bastian}, {Bressert}, {Molinari},
  {Battersby}, \& {Walsh}}]{LongmoreEtAl2013b}
{Longmore}, S.~N., {Kruijssen}, J.~M.~D., {Bally}, J., {et~al.}
  2013{\natexlab{a}}, \mnras, 433, L15

\bibitem[{{Longmore} {et~al.}(2013{\natexlab{b}}){Longmore}, {Bally}, {Testi},
  {Purcell}, {Walsh}, {Bressert}, {Pestalozzi}, {Molinari}, {Ott}, {Cortese},
  {Battersby}, {Murray}, {Lee}, {Kruijssen}, {Schisano}, \&
  {Elia}}]{LongmoreEtAl2013a}
{Longmore}, S.~N., {Bally}, J., {Testi}, L., {et~al.} 2013{\natexlab{b}},
  \mnras, 429, 987

\bibitem[{{Mac Low}(1999)}]{MacLow1999}
{Mac Low}, M.-M. 1999, \apj, 524, 169

\bibitem[{{Mac Low} {et~al.}(1998){Mac Low}, {Klessen}, {Burkert}, \&
  {Smith}}]{MacLowEtAl1998}
{Mac Low}, M.-M., {Klessen}, R.~S., {Burkert}, A., \& {Smith}, M.~D. 1998,
  PhRvL, 80, 2754

\bibitem[{{Malinen} {et~al.}(2012){Malinen}, {Juvela}, {Rawlings},
  {Ward-Thompson}, {Palmeirim}, \& {Andr{\'e}}}]{MalinenEtAl2012}
{Malinen}, J., {Juvela}, M., {Rawlings}, M.~G., {et~al.} 2012, \aap, 544, A50

\bibitem[{{Malkin}(2013)}]{Malkin2013}
{Malkin}, Z.~M. 2013, Astronomy Reports, 57, 128

\bibitem[{{Marsh} {et~al.}(2016){Marsh}, {Ragan}, {Whitworth}, \&
  {Clark}}]{MarshEtAl2016}
{Marsh}, K.~A., {Ragan}, S.~E., {Whitworth}, A.~P., \& {Clark}, P.~C. 2016,
  \mnras, accepted (arXiv:1604.07609)

\bibitem[{{Martig} {et~al.}(2013){Martig}, {Crocker}, {Bournaud}, {Emsellem},
  {Gabor}, {Alatalo}, {Blitz}, {Bois}, {Bureau}, {Cappellari}, {Davies},
  {Davis}, {Dekel}, {de Zeeuw}, {Duc}, {Falc{\'o}n-Barroso}, {Khochfar},
  {Krajnovi{\'c}}, {Kuntschner}, {Morganti}, {McDermid}, {Naab}, {Oosterloo},
  {Sarzi}, {Scott}, {Serra}, {Griffin}, {Teyssier}, {Weijmans}, \&
  {Young}}]{MartigEtAl2013}
{Martig}, M., {Crocker}, A.~F., {Bournaud}, F., {et~al.} 2013, \mnras, 432,
  1914

\bibitem[{{Matzner} \& {McKee}(2000)}]{MatznerMcKee2000}
{Matzner}, C.~D., \& {McKee}, C.~F. 2000, \apj, 545, 364

\bibitem[{{McKee}(1989)}]{McKee1989}
{McKee}, C.~F. 1989, \apj, 345, 782

\bibitem[{{McKee} \& {Ostriker}(2007)}]{McKeeOstriker2007}
{McKee}, C.~F., \& {Ostriker}, E.~C. 2007, \araa, 45, 565

\bibitem[{{McMullin} {et~al.}(2007){McMullin}, {Waters}, {Schiebel}, {Young},
  \& {Golap}}]{casa}
{McMullin}, J.~P., {Waters}, B., {Schiebel}, D., {Young}, W., \& {Golap}, K.
  2007, in Astronomical Society of the Pacific Conference Series, Vol. 376,
  Astronomical Data Analysis Software and Systems XVI, ed. R.~A. {Shaw},
  F.~{Hill}, \& D.~J. {Bell}, 127

\bibitem[{{Mee} \& {Brandenburg}(2006)}]{MeeBrandenburg2006}
{Mee}, A.~J., \& {Brandenburg}, A. 2006, \mnras, 370, 415

\bibitem[{{Mills} {et~al.}(2015){Mills}, {Butterfield}, {Ludovici}, {Lang},
  {Ott}, {Morris}, \& {Schmitz}}]{MillsEtAl2015}
{Mills}, E.~A.~C., {Butterfield}, N., {Ludovici}, D.~A., {et~al.} 2015, \apj,
  805, 72

\bibitem[{{Mills} \& {Morris}(2013)}]{MillsMorris2013}
{Mills}, E.~A.~C., \& {Morris}, M.~R. 2013, \apj, 772, 105

\bibitem[{{Moeckel} \& {Burkert}(2015)}]{MoeckelBurkert2015}
{Moeckel}, N., \& {Burkert}, A. 2015, \apj, 807, 67

\bibitem[{{Molina} {et~al.}(2012){Molina}, {Glover}, {Federrath}, \&
  {Klessen}}]{MolinaEtAl2012}
{Molina}, F.~Z., {Glover}, S.~C.~O., {Federrath}, C., \& {Klessen}, R.~S. 2012,
  \mnras, 423, 2680

\bibitem[{{Molinari} {et~al.}(2010){Molinari}, {Swinyard}, {Bally}, {Barlow},
  {Bernard}, {Martin}, {Moore}, {Noriega-Crespo}, {Plume}, {Testi}, {Zavagno},
  {Abergel}, {Ali}, {Andr{\'e}}, {Baluteau}, {Benedettini}, {Bern{\'e}},
  {Billot}, {Blommaert}, {Bontemps}, {Boulanger}, {Brand}, {Brunt}, {Burton},
  {Campeggio}, {Carey}, {Caselli}, {Cesaroni}, {Cernicharo}, {Chakrabarti},
  {Chrysostomou}, {Codella}, {Cohen}, {Compiegne}, {Davis}, {de Bernardis}, {de
  Gasperis}, {Di Francesco}, {di Giorgio}, {Elia}, {Faustini}, {Fischera},
  {Fukui}, {Fuller}, {Ganga}, {Garcia-Lario}, {Giard}, {Giardino}, {Glenn},
  {Goldsmith}, {Griffin}, {Hoare}, {Huang}, {Jiang}, {Joblin}, {Joncas},
  {Juvela}, {Kirk}, {Lagache}, {Li}, {Lim}, {Lord}, {Lucas}, {Maiolo},
  {Marengo}, {Marshall}, {Masi}, {Massi}, {Matsuura}, {Meny}, {Minier},
  {Miville-Desch{\^e}nes}, {Montier}, {Motte}, {M{\"u}ller}, {Natoli}, {Neves},
  {Olmi}, {Paladini}, {Paradis}, {Pestalozzi}, {Pezzuto}, {Piacentini},
  {Pomar{\`e}s}, {Popescu}, {Reach}, {Richer}, {Ristorcelli}, {Roy}, {Royer},
  {Russeil}, {Saraceno}, {Sauvage}, {Schilke}, {Schneider-Bontemps},
  {Schuller}, {Schultz}, {Shepherd}, {Sibthorpe}, {Smith}, {Smith},
  {Spinoglio}, {Stamatellos}, {Strafella}, {Stringfellow}, {Sturm}, {Taylor},
  {Thompson}, {Tuffs}, {Umana}, {Valenziano}, {Vavrek}, {Viti}, {Waelkens},
  {Ward-Thompson}, {White}, {Wyrowski}, {Yorke}, \& {Zhang}}]{MolinariEtAl2010}
{Molinari}, S., {Swinyard}, B., {Bally}, J., {et~al.} 2010, \pasp, 122, 314

\bibitem[{{Molinari} {et~al.}(2011){Molinari}, {Bally}, {Noriega-Crespo},
  {Compi{\`e}gne}, {Bernard}, {Paradis}, {Martin}, {Testi}, {Barlow}, {Moore},
  {Plume}, {Swinyard}, {Zavagno}, {Calzoletti}, {Di Giorgio}, {Elia},
  {Faustini}, {Natoli}, {Pestalozzi}, {Pezzuto}, {Piacentini}, {Polenta},
  {Polychroni}, {Schisano}, {Traficante}, {Veneziani}, {Battersby}, {Burton},
  {Carey}, {Fukui}, {Li}, {Lord}, {Morgan}, {Motte}, {Schuller},
  {Stringfellow}, {Tan}, {Thompson}, {Ward-Thompson}, {White}, \&
  {Umana}}]{MolinariEtAl2011}
{Molinari}, S., {Bally}, J., {Noriega-Crespo}, A., {et~al.} 2011, \apjl, 735,
  L33

\bibitem[{{Myers}(2011)}]{Myers2011}
{Myers}, P.~C. 2011, \apj, 735, 82

\bibitem[{{Nakamura} \& {Li}(2008)}]{NakamuraLi2008}
{Nakamura}, F., \& {Li}, Z. 2008, \apj, 687, 354

\bibitem[{{Nolan} {et~al.}(2015){Nolan}, {Federrath}, \&
  {Sutherland}}]{NolanFederrathSutherland2015}
{Nolan}, C.~A., {Federrath}, C., \& {Sutherland}, R.~S. 2015, \mnras, 451, 1380

\bibitem[{{Norman} \& {Silk}(1980)}]{NormanSilk1980}
{Norman}, C., \& {Silk}, J. 1980, \apj, 238, 158

\bibitem[{{Nutter} {et~al.}(2008){Nutter}, {Kirk}, {Stamatellos}, \&
  {Ward-Thompson}}]{NutterEtAl2008}
{Nutter}, D., {Kirk}, J.~M., {Stamatellos}, D., \& {Ward-Thompson}, D. 2008,
  \mnras, 384, 755

\bibitem[{{Offner} \& {Arce}(2014)}]{OffnerArce2014}
{Offner}, S.~S.~R., \& {Arce}, H.~G. 2014, \apj, 784, 61

\bibitem[{{Ossenkopf} {et~al.}(2008){Ossenkopf}, {Krips}, \&
  {Stutzki}}]{OssenkopfKripsStutzki2008a}
{Ossenkopf}, V., {Krips}, M., \& {Stutzki}, J. 2008, \aap, 485, 917

\bibitem[{{Ostriker}(1964)}]{Ostriker1964}
{Ostriker}, J. 1964, \apj, 140, 1056

\bibitem[{{Padoan} {et~al.}(2014){Padoan}, {Federrath}, {Chabrier}, {Evans},
  {Johnstone}, {J{\o}rgensen}, {McKee}, \& {Nordlund}}]{PadoanEtAl2014}
{Padoan}, P., {Federrath}, C., {Chabrier}, G., {et~al.} 2014, Protostars and
  Planets VI, 77

\bibitem[{{Padoan} {et~al.}(1997{\natexlab{a}}){Padoan}, {Jones}, \&
  {Nordlund}}]{PadoanJonesNordlund1997}
{Padoan}, P., {Jones}, B.~J.~T., \& {Nordlund}, A.~P. 1997{\natexlab{a}}, \apj,
  474, 730

\bibitem[{{Padoan} \& {Nordlund}(2002)}]{PadoanNordlund2002}
{Padoan}, P., \& {Nordlund}, {\AA}. 2002, \apj, 576, 870

\bibitem[{{Padoan} \& {Nordlund}(2011)}]{PadoanNordlund2011}
---. 2011, \apj, 730, 40

\bibitem[{{Padoan} {et~al.}(1997{\natexlab{b}}){Padoan}, {Nordlund}, \&
  {Jones}}]{PadoanNordlundJones1997}
{Padoan}, P., {Nordlund}, {\AA}., \& {Jones}, B.~J.~T. 1997{\natexlab{b}},
  \mnras, 288, 145

\bibitem[{{Padoan} {et~al.}(2016){Padoan}, {Pan}, {Haugb{\o}lle}, \&
  {Nordlund}}]{PadoanEtAl2016}
{Padoan}, P., {Pan}, L., {Haugb{\o}lle}, T., \& {Nordlund}, {\AA}. 2016, \apj,
  822, 11

\bibitem[{{Palmeirim} {et~al.}(2013){Palmeirim}, {Andr{\'e}}, {Kirk},
  {Ward-Thompson}, {Arzoumanian}, {K{\"o}nyves}, {Didelon}, {Schneider},
  {Benedettini}, {Bontemps}, {Di Francesco}, {Elia}, {Griffin}, {Hennemann},
  {Hill}, {Martin}, {Men'shchikov}, {Molinari}, {Motte}, {Nguyen Luong},
  {Nutter}, {Peretto}, {Pezzuto}, {Roy}, {Rygl}, {Spinoglio}, \&
  {White}}]{PalmeirimEtAl2013}
{Palmeirim}, P., {Andr{\'e}}, P., {Kirk}, J., {et~al.} 2013, \aap, 550, A38

\bibitem[{{Pan} {et~al.}(2015){Pan}, {Padoan}, {Haugbolle}, \&
  {Nordlund}}]{PanEtAl2016}
{Pan}, L., {Padoan}, P., {Haugbolle}, T., \& {Nordlund}, A. 2015, \apj,
  accepted (arXiv:1510.04742)

\bibitem[{{Passot} \& {V{\'a}zquez-Semadeni}(1998)}]{PassotVazquez1998}
{Passot}, T., \& {V{\'a}zquez-Semadeni}, E. 1998, PhRvE, 58, 4501

\bibitem[{{Peters} {et~al.}(2011){Peters}, {Banerjee}, {Klessen}, \& {Mac
  Low}}]{PetersEtAl2011}
{Peters}, T., {Banerjee}, R., {Klessen}, R.~S., \& {Mac Low}, M. 2011, \apj,
  729, 72

\bibitem[{{Pillai} {et~al.}(2015){Pillai}, {Kauffmann}, {Tan}, {Goldsmith},
  {Carey}, \& {Menten}}]{PillaiEtAl2015}
{Pillai}, T., {Kauffmann}, J., {Tan}, J.~C., {et~al.} 2015, \apj, 799, 74

\bibitem[{{Pineda} {et~al.}(2011){Pineda}, {Goodman}, {Arce}, {Caselli},
  {Longmore}, \& {Corder}}]{PinedaEtAl2011}
{Pineda}, J.~E., {Goodman}, A.~A., {Arce}, H.~G., {et~al.} 2011, \apjl, 739, L2

\bibitem[{{Piontek} \& {Ostriker}(2007)}]{PiontekOstriker2007}
{Piontek}, R.~A., \& {Ostriker}, E.~C. 2007, \apj, 663, 183

\bibitem[{{Planck Collaboration} {et~al.}(2014){Planck Collaboration}, {Adam},
  {Ade}, {Aghanim}, {Alves}, {Arnaud}, {Arzoumanian}, {Ashdown}, {Aumont},
  {Baccigalupi}, {Banday}, {Barreiro}, {Bartolo}, {Battaner}, {Benabed},
  {Benoit-L{\'e}vy}, {Bernard}, {Bersanelli}, {Bielewicz}, {Bonaldi},
  {Bonavera}, {Bond}, {Borrill}, {Bouchet}, {Boulanger}, {Bracco}, {Burigana},
  {Butler}, {Calabrese}, {Cardoso}, {Catalano}, {Chamballu}, {Chiang},
  {Christensen}, {Colombi}, {Colombo}, {Combet}, {Couchot}, {Crill}, {Curto},
  {Cuttaia}, {Danese}, {Davies}, {Davis}, {de Bernardis}, {de Rosa}, {de
  Zotti}, {Delabrouille}, {Dickinson}, {Diego}, {Dole}, {Donzelli}, {Dor{\'e}},
  {Douspis}, {Ducout}, {Dupac}, {Efstathiou}, {Elsner}, {En{\ss}lin},
  {Eriksen}, {Falgarone}, {Ferri{\`e}re}, {Finelli}, {Forni}, {Frailis},
  {Fraisse}, {Franceschi}, {Frejsel}, {Galeotta}, {Galli}, {Ganga}, {Ghosh},
  {Giard}, {Gjerl{\o}w}, {Gonz{\'a}lez-Nuevo}, {G{\'o}rski}, {Gregorio},
  {Gruppuso}, {Guillet}, {Hansen}, {Hanson}, {Harrison}, {Henrot-Versill{\'e}},
  {Hern{\'a}ndez-Monteagudo}, {Herranz}, {Hildebrandt}, {Hivon}, {Holmes},
  {Hovest}, {Huffenberger}, {Hurier}, {Jaffe}, {Jaffe}, {Jones},
  {Keih{\"a}nen}, {Keskitalo}, {Kisner}, {Kneissl}, {Knoche}, {Kunz},
  {Kurki-Suonio}, {Lagache}, {Lamarre}, {Lasenby}, {Lattanzi}, {Lawrence},
  {Leonardi}, {Levrier}, {Liguori}, {Lilje}, {Linden-V{\o}rnle},
  {L{\'o}pez-Caniego}, {Lubin}, {Mac{\'{\i}}as-P{\'e}rez}, {Maffei}, {Maino},
  {Mandolesi}, {Maris}, {Marshall}, {Martin}, {Mart{\'{\i}}nez-Gonz{\'a}lez},
  {Masi}, {Matarrese}, {Mazzotta}, {Melchiorri}, {Mendes}, {Mennella},
  {Migliaccio}, {Miville-Desch{\^e}nes}, {Moneti}, {Montier}, {Morgante},
  {Mortlock}, {Munshi}, {Murphy}, {Naselsky}, {Natoli}, {N{\o}rgaard-Nielsen},
  {Noviello}, {Novikov}, {Novikov}, {Oppermann}, {Oxborrow}, {Pagano}, {Pajot},
  {Paoletti}, {Pasian}, {Perdereau}, {Perotto}, {Perrotta}, {Pettorino},
  {Piacentini}, {Piat}, {Plaszczynski}, {Pointecouteau}, {Polenta}, {Ponthieu},
  {Popa}, {Pratt}, {Prunet}, {Puget}, {Rachen}, {Reach}, {Reinecke},
  {Remazeilles}, {Renault}, {Ristorcelli}, {Rocha}, {Roudier},
  {Rubi{\~n}o-Mart{\'{\i}}n}, {Rusholme}, {Sandri}, {Santos}, {Savini},
  {Scott}, {Soler}, {Spencer}, {Stolyarov}, {Sudiwala}, {Sunyaev}, {Sutton},
  {Suur-Uski}, {Sygnet}, {Tauber}, {Terenzi}, {Toffolatti}, {Tomasi},
  {Tristram}, {Tucci}, {Umana}, {Valenziano}, {Valiviita}, {Van Tent},
  {Vielva}, {Villa}, {Wade}, {Wandelt}, {Wehus}, {Wiesemeyer}, {Yvon},
  {Zacchei}, \& {Zonca}}]{PlanckMagneticFilaments2014}
{Planck Collaboration}, {Adam}, R., {Ade}, P.~A.~R., {et~al.} 2014, \aap,
  submitted (arXiv:1409.6728)

\bibitem[{{Planck Collaboration} {et~al.}(2015{\natexlab{a}}){Planck
  Collaboration}, {Ade}, {Aghanim}, {Alves}, {Arnaud}, {Arzoumanian},
  {Ashdown}, {Aumont}, {Baccigalupi}, {Banday}, {Barreiro}, {Bartolo},
  {Battaner}, {Benabed}, {Beno{\^i}t}, {Benoit-L{\'e}vy}, {Bernard},
  {Bersanelli}, {Bielewicz}, {Bock}, {Bonavera}, {Bond}, {Borrill}, {Bouchet},
  {Boulanger}, {Bracco}, {Burigana}, {Calabrese}, {Cardoso}, {Catalano},
  {Chiang}, {Christensen}, {Colombo}, {Combet}, {Couchot}, {Crill}, {Curto},
  {Cuttaia}, {Danese}, {Davies}, {Davis}, {de Bernardis}, {de Rosa}, {de
  Zotti}, {Delabrouille}, {Dickinson}, {Diego}, {Dole}, {Donzelli}, {Dor{\'e}},
  {Douspis}, {Ducout}, {Dupac}, {Efstathiou}, {Elsner}, {En{\ss}lin},
  {Eriksen}, {Falgarone}, {Ferri{\`e}re}, {Finelli}, {Forni}, {Frailis},
  {Fraisse}, {Franceschi}, {Frejsel}, {Galeotta}, {Galli}, {Ganga}, {Ghosh},
  {Giard}, {Gjerl{\o}w}, {Gonz{\'a}lez-Nuevo}, {G{\'o}rski}, {Gregorio},
  {Gruppuso}, {Gudmundsson}, {Guillet}, {Harrison}, {Helou},
  {Henrot-Versill{\'e}}, {Hern{\'a}ndez-Monteagudo}, {Herranz}, {Hildebrandt},
  {Hivon}, {Holmes}, {Hornstrup}, {Huffenberger}, {Hurier}, {Jaffe}, {Jaffe},
  {Jones}, {Juvela}, {Keih{\"a}nen}, {Keskitalo}, {Kisner}, {Knoche}, {Kunz},
  {Kurki-Suonio}, {Lagache}, {Lamarre}, {Lasenby}, {Lattanzi}, {Lawrence},
  {Leonardi}, {Levrier}, {Liguori}, {Lilje}, {Linden-V{\o}rnle},
  {L{\'o}pez-Caniego}, {Lubin}, {Mac{\'{\i}}as-P{\'e}rez}, {Maino},
  {Mandolesi}, {Mangilli}, {Maris}, {Martin}, {Mart{\'{\i}}nez-Gonz{\'a}lez},
  {Masi}, {Matarrese}, {Melchiorri}, {Mendes}, {Mennella}, {Migliaccio},
  {Miville-Desch{\^e}nes}, {Moneti}, {Montier}, {Morgante}, {Mortlock},
  {Munshi}, {Murphy}, {Naselsky}, {Nati}, {Netterfield}, {Noviello}, {Novikov},
  {Novikov}, {Oppermann}, {Oxborrow}, {Pagano}, {Pajot}, {Paladini},
  {Paoletti}, {Pasian}, {Perotto}, {Pettorino}, {Piacentini}, {Piat},
  {Pierpaoli}, {Pietrobon}, {Plaszczynski}, {Pointecouteau}, {Polenta},
  {Ponthieu}, {Pratt}, {Prunet}, {Puget}, {Rachen}, {Reinecke}, {Remazeilles},
  {Renault}, {Renzi}, {Ristorcelli}, {Rocha}, {Rossetti}, {Roudier},
  {Rubi{\~n}o-Mart{\'{\i}}n}, {Rusholme}, {Sandri}, {Santos}, {Savelainen},
  {Savini}, {Scott}, {Soler}, {Stolyarov}, {Sudiwala}, {Sutton}, {Suur-Uski},
  {Sygnet}, {Tauber}, {Terenzi}, {Toffolatti}, {Tomasi}, {Tristram}, {Tucci},
  {Umana}, {Valenziano}, {Valiviita}, {Van Tent}, {Vielva}, {Villa}, {Wade},
  {Wandelt}, {Wehus}, {Ysard}, {Yvon}, \&
  {Zonca}}]{PlanckMagneticFilaments2015a}
{Planck Collaboration}, {Ade}, P.~A.~R., {Aghanim}, N., {et~al.}
  2015{\natexlab{a}}, \aap, accepted (arXiv:1502.04123)

\bibitem[{{Planck Collaboration} {et~al.}(2015{\natexlab{b}}){Planck
  Collaboration}, {Ade}, {Aghanim}, {Arnaud}, {Ashdown}, {Aumont},
  {Baccigalupi}, {Banday}, {Barreiro}, {Bartolo}, {Battaner}, {Benabed},
  {Benoit-L{\'e}vy}, {Bernard}, {Bersanelli}, {Bielewicz}, {Bonaldi},
  {Bonavera}, {Bond}, {Borrill}, {Bouchet}, {Boulanger}, {Bracco}, {Burigana},
  {Calabrese}, {Cardoso}, {Catalano}, {Chamballu}, {Chary}, {Chiang},
  {Christensen}, {Colombo}, {Combet}, {Crill}, {Curto}, {Cuttaia}, {Danese},
  {Davies}, {Davis}, {de Bernardis}, {de Rosa}, {de Zotti}, {Delabrouille},
  {Delouis}, {Dickinson}, {Diego}, {Dole}, {Donzelli}, {Dor{\'e}}, {Douspis},
  {Dunkley}, {Dupac}, {Efstathiou}, {Elsner}, {En{\ss}lin}, {Eriksen},
  {Falgarone}, {Ferri{\`e}re}, {Finelli}, {Forni}, {Frailis}, {Fraisse},
  {Franceschi}, {Frolov}, {Galeotta}, {Galli}, {Ganga}, {Ghosh}, {Giard},
  {Gjerl{\o}w}, {Gonz{\'a}lez-Nuevo}, {G{\'o}rski}, {Gruppuso}, {Guillet},
  {Hansen}, {Harrison}, {Helou}, {Hern{\'a}ndez-Monteagudo}, {Herranz},
  {Hildebrandt}, {Hivon}, {Hornstrup}, {Hovest}, {Huang}, {Huffenberger},
  {Hurier}, {Jaffe}, {Jones}, {Juvela}, {Keih{\"a}nen}, {Keskitalo}, {Kisner},
  {Kneissl}, {Knoche}, {Kunz}, {Kurki-Suonio}, {Lamarre}, {Lasenby},
  {Lattanzi}, {Lawrence}, {Leonardi}, {Le{\'o}n-Tavares}, {Levrier}, {Liguori},
  {Lilje}, {Linden-V{\o}rnle}, {L{\'o}pez-Caniego}, {Lubin},
  {Mac{\'{\i}}as-P{\'e}rez}, {Maffei}, {Maino}, {Mandolesi}, {Maris}, {Martin},
  {Mart{\'{\i}}nez-Gonz{\'a}lez}, {Masi}, {Matarrese}, {McGehee}, {Melchiorri},
  {Mennella}, {Migliaccio}, {Miville-Desch{\^e}nes}, {Moneti}, {Montier},
  {Morgante}, {Mortlock}, {Munshi}, {Murphy}, {Naselsky}, {Nati}, {Natoli},
  {Novikov}, {Novikov}, {Oppermann}, {Oxborrow}, {Pagano}, {Pajot}, {Paoletti},
  {Pasian}, {Perdereau}, {Pettorino}, {Piacentini}, {Piat}, {Pierpaoli},
  {Plaszczynski}, {Pointecouteau}, {Polenta}, {Ponthieu}, {Pratt}, {Prunet},
  {Puget}, {Rachen}, {Reach}, {Rebolo}, {Reinecke}, {Remazeilles}, {Renault},
  {Renzi}, {Ristorcelli}, {Rocha}, {Rosset}, {Rossetti}, {Roudier},
  {Rubi{\~n}o-Mart{\'{\i}}n}, {Rusholme}, {Sandri}, {Santos}, {Savelainen},
  {Savini}, {Scott}, {Serra}, {Soler}, {Stolyarov}, {Sudiwala}, {Sunyaev},
  {Suur-Uski}, {Sygnet}, {Tauber}, {Terenzi}, {Toffolatti}, {Tomasi},
  {Tristram}, {Tucci}, {Umana}, {Valenziano}, {Valiviita}, {Van Tent},
  {Vielva}, {Villa}, {Wade}, {Wandelt}, {Wehus}, {Yvon}, {Zacchei}, \&
  {Zonca}}]{PlanckMagneticFilaments2015b}
---. 2015{\natexlab{b}}, \aap, accepted (arXiv:1505.02779)

\bibitem[{{Plunkett} {et~al.}(2015){Plunkett}, {Arce}, {Corder}, {Dunham},
  {Garay}, \& {Mardones}}]{PlunkettEtAl2015}
{Plunkett}, A.~L., {Arce}, H.~G., {Corder}, S.~A., {et~al.} 2015, \apj, 803, 22

\bibitem[{{Plunkett} {et~al.}(2013){Plunkett}, {Arce}, {Corder}, {Mardones},
  {Sargent}, \& {Schnee}}]{PlunkettEtAl2013}
---. 2013, \apj, 774, 22

\bibitem[{{Polychroni} {et~al.}(2013){Polychroni}, {Schisano}, {Elia}, {Roy},
  {Molinari}, {Martin}, {Andr{\'e}}, {Turrini}, {Rygl}, {Di Francesco},
  {Benedettini}, {Busquet}, {di Giorgio}, {Pestalozzi}, {Pezzuto},
  {Arzoumanian}, {Bontemps}, {Hennemann}, {Hill}, {K{\"o}nyves},
  {Men'shchikov}, {Motte}, {Nguyen-Luong}, {Peretto}, {Schneider}, \&
  {White}}]{PolychroniEtAl2013}
{Polychroni}, D., {Schisano}, E., {Elia}, D., {et~al.} 2013, \apjl, 777, L33

\bibitem[{{Price} {et~al.}(2011){Price}, {Federrath}, \&
  {Brunt}}]{PriceFederrathBrunt2011}
{Price}, D.~J., {Federrath}, C., \& {Brunt}, C.~M. 2011, \apjl, 727, L21

\bibitem[{{Rathborne} {et~al.}(2014{\natexlab{a}}){Rathborne}, {Longmore},
  {Jackson}, {Foster}, {Contreras}, {Garay}, {Testi}, {Alves}, {Bally},
  {Bastian}, {Kruijssen}, \& {Bressert}}]{RathborneEtAl2014b}
{Rathborne}, J.~M., {Longmore}, S.~N., {Jackson}, J.~M., {et~al.}
  2014{\natexlab{a}}, \apj, 786, 140

\bibitem[{{Rathborne} {et~al.}(2014{\natexlab{b}}){Rathborne}, {Longmore},
  {Jackson}, {Kruijssen}, {Alves}, {Bally}, {Bastian}, {Contreras}, {Foster},
  {Garay}, {Testi}, \& {Walsh}}]{RathborneEtAl2014}
---. 2014{\natexlab{b}}, \apjl, 795, L25

\bibitem[{{Rathborne} {et~al.}(2015){Rathborne}, {Longmore}, {Jackson},
  {Alves}, {Bally}, {Bastian}, {Contreras}, {Foster}, {Garay}, {Kruijssen},
  {Testi}, \& {Walsh}}]{RathborneEtAl2015}
---. 2015, \apj, 802, 125

\bibitem[{{Reid} {et~al.}(2014){Reid}, {Menten}, {Brunthaler}, {Zheng}, {Dame},
  {Xu}, {Wu}, {Zhang}, {Sanna}, {Sato}, {Hachisuka}, {Choi}, {Immer},
  {Moscadelli}, {Rygl}, \& {Bartkiewicz}}]{ReidEtAl2014}
{Reid}, M.~J., {Menten}, K.~M., {Brunthaler}, A., {et~al.} 2014, \apj, 783, 130

\bibitem[{{Robertson} \& {Goldreich}(2012)}]{RobertsonGoldreich2012}
{Robertson}, B., \& {Goldreich}, P. 2012, \apjl, 750, L31

\bibitem[{{Roy} {et~al.}(2015){Roy}, {Andr{\'e}}, {Arzoumanian}, {Peretto},
  {Palmeirim}, {K{\"o}nyves}, {Schneider}, {Benedettini}, {Di Francesco},
  {Elia}, {Hill}, {Ladjelate}, {Louvet}, {Motte}, {Pezzuto}, {Schisano},
  {Shimajiri}, {Spinoglio}, {Ward-Thompson}, \& {White}}]{RoyEtAl2015}
{Roy}, A., {Andr{\'e}}, P., {Arzoumanian}, D., {et~al.} 2015, \aap, 584, A111

\bibitem[{{Salim} {et~al.}(2015){Salim}, {Federrath}, \&
  {Kewley}}]{SalimFederrathKewley2015}
{Salim}, D.~M., {Federrath}, C., \& {Kewley}, L.~J. 2015, \apjl, 806, L36

\bibitem[{{Salji} {et~al.}(2015){Salji}, {Richer}, {Buckle}, {Francesco},
  {Hatchell}, {Hogerheijde}, {Johnstone}, {Kirk}, {Ward-Thompson}, \& {JCMT GBS
  Consortium}}]{SaljiEtAl2015}
{Salji}, C.~J., {Richer}, J.~S., {Buckle}, J.~V., {et~al.} 2015, \mnras, 449,
  1782

\bibitem[{{Sault} {et~al.}(1995){Sault}, {Teuben}, \& {Wright}}]{Miriad}
{Sault}, R.~J., {Teuben}, P.~J., \& {Wright}, M.~C.~H. 1995, in Astronomical
  Society of the Pacific Conference Series, Vol.~77, Astronomical Data Analysis
  Software and Systems IV, ed. R.~A. {Shaw}, H.~E. {Payne}, \& J.~J.~E.
  {Hayes}, 433

\bibitem[{{Scalo} \& {Pumphrey}(1982)}]{ScaloPumphrey1982}
{Scalo}, J.~M., \& {Pumphrey}, W.~A. 1982, \apjl, 258, L29

\bibitem[{{Schmidt} {et~al.}(2009){Schmidt}, {Federrath}, {Hupp}, {Kern}, \&
  {Niemeyer}}]{SchmidtEtAl2009}
{Schmidt}, W., {Federrath}, C., {Hupp}, M., {Kern}, S., \& {Niemeyer}, J.~C.
  2009, \aap, 494, 127

\bibitem[{{Schmidt} {et~al.}(2008){Schmidt}, {Federrath}, \&
  {Klessen}}]{SchmidtFederrathKlessen2008}
{Schmidt}, W., {Federrath}, C., \& {Klessen}, R. 2008, PhRvL, 101, 194505

\bibitem[{{Schneider} {et~al.}(2012){Schneider}, {Csengeri}, {Hennemann},
  {Motte}, {Didelon}, {Federrath}, {Bontemps}, {Di Francesco}, {Arzoumanian},
  {Minier}, {Andr{\'e}}, {Hill}, {Zavagno}, {Nguyen-Luong}, {Attard},
  {Bernard}, {Elia}, {Fallscheer}, {Griffin}, {Kirk}, {Klessen}, {K{\"o}nyves},
  {Martin}, {Men'shchikov}, {Palmeirim}, {Peretto}, {Pestalozzi}, {Russeil},
  {Sadavoy}, {Sousbie}, {Testi}, {Tremblin}, {Ward-Thompson}, \&
  {White}}]{SchneiderEtAl2012}
{Schneider}, N., {Csengeri}, T., {Hennemann}, M., {et~al.} 2012, \aap, 540, L11

\bibitem[{{Schneider} {et~al.}(2013){Schneider}, {Andr{\'e}}, {K{\"o}nyves},
  {Bontemps}, {Motte}, {Federrath}, {Ward-Thompson}, {Arzoumanian},
  {Benedettini}, {Bressert}, {Didelon}, {Di Francesco}, {Griffin}, {Hennemann},
  {Hill}, {Palmeirim}, {Pezzuto}, {Peretto}, {Roy}, {Rygl}, {Spinoglio}, \&
  {White}}]{SchneiderEtAl2013}
{Schneider}, N., {Andr{\'e}}, P., {K{\"o}nyves}, V., {et~al.} 2013, \apjl, 766,
  L17

\bibitem[{{Schneider} {et~al.}(2015){Schneider}, {Ossenkopf}, {Csengeri},
  {Klessen}, {Federrath}, {Tremblin}, {Girichidis}, {Bontemps}, \&
  {Andr{\'e}}}]{SchneiderEtAl2015}
{Schneider}, N., {Ossenkopf}, V., {Csengeri}, T., {et~al.} 2015, \aap, 575, A79

\bibitem[{{Schneider} \& {Elmegreen}(1979)}]{SchneiderElmegreen1979}
{Schneider}, S., \& {Elmegreen}, B.~G. 1979, \apjs, 41, 87

\bibitem[{{Seifried} \& {Walch}(2015)}]{SeifriedWalch2015}
{Seifried}, D., \& {Walch}, S. 2015, \mnras, 452, 2410

\bibitem[{{Smith} {et~al.}(2014){Smith}, {Glover}, \&
  {Klessen}}]{SmithGloverKlessen2014}
{Smith}, R.~J., {Glover}, S.~C.~O., \& {Klessen}, R.~S. 2014, \mnras, 445, 2900

\bibitem[{{Smith} {et~al.}(2016){Smith}, {Glover}, {Klessen}, \&
  {Fuller}}]{SmithEtAl2016}
{Smith}, R.~J., {Glover}, S.~C.~O., {Klessen}, R.~S., \& {Fuller}, G.~A. 2016,
  \mnras, 455, 3640

\bibitem[{{Sofue} {et~al.}(1987){Sofue}, {Reich}, {Inoue}, \&
  {Seiradakis}}]{SofueEtAl1987}
{Sofue}, Y., {Reich}, W., {Inoue}, M., \& {Seiradakis}, J.~H. 1987, \pasj, 39,
  95

\bibitem[{{Sousbie}(2011)}]{Sousbie2011}
{Sousbie}, T. 2011, \mnras, 414, 350

\bibitem[{{Sousbie} {et~al.}(2011){Sousbie}, {Pichon}, \&
  {Kawahara}}]{SousbieEtAl2011}
{Sousbie}, T., {Pichon}, C., \& {Kawahara}, H. 2011, \mnras, 414, 384

\bibitem[{{Stone} {et~al.}(1998){Stone}, {Ostriker}, \&
  {Gammie}}]{StoneOstrikerGammie1998}
{Stone}, J.~M., {Ostriker}, E.~C., \& {Gammie}, C.~F. 1998, \apjl, 508, L99

\bibitem[{{Sugitani} {et~al.}(2011){Sugitani}, {Nakamura}, {Watanabe},
  {Tamura}, {Nishiyama}, {Nagayama}, {Kandori}, {Nagata}, {Sato}, {Gutermuth},
  {Wilson}, \& {Kawabe}}]{SugitaniEtAl2011}
{Sugitani}, K., {Nakamura}, F., {Watanabe}, M., {et~al.} 2011, \apj, 734, 63

\bibitem[{{Sun} \& {Takayama}(2003)}]{SunTakayama2003}
{Sun}, M., \& {Takayama}, K. 2003, JFM, 478, 237

\bibitem[{{Sur} {et~al.}(2010){Sur}, {Schlei\-cher}, {Banerjee}, {Federrath},
  \& {Klessen}}]{SurEtAl2010}
{Sur}, S., {Schlei\-cher}, D.~R.~G., {Banerjee}, R., {Federrath}, C., \&
  {Klessen}, R.~S. 2010, \apjl, 721, L134

\bibitem[{{Tamburro} {et~al.}(2009){Tamburro}, {Rix}, {Leroy}, {Low}, {Walter},
  {Kennicutt}, {Brinks}, \& {de Blok}}]{TamburroEtAl2009}
{Tamburro}, D., {Rix}, H.-W., {Leroy}, A.~K., {et~al.} 2009, \aj, 137, 4424

\bibitem[{{Tomisaka}(2014)}]{Tomisaka2014}
{Tomisaka}, K. 2014, \apj, 785, 24

\bibitem[{{Tsuboi} {et~al.}(1986){Tsuboi}, {Inoue}, {Handa}, {Tabara}, {Kato},
  {Sofue}, \& {Kaifu}}]{TsuboiEtAl1986}
{Tsuboi}, M., {Inoue}, M., {Handa}, T., {et~al.} 1986, \aj, 92, 818

\bibitem[{{V{\'a}zquez-Semadeni} {et~al.}(2003){V{\'a}zquez-Semadeni},
  {Ballesteros-Paredes}, \& {Klessen}}]{VazquezBallesterosKlessen2003}
{V{\'a}zquez-Semadeni}, E., {Ballesteros-Paredes}, J., \& {Klessen}, R.~S.
  2003, \apjl, 585, L131

\bibitem[{{Vazquez-Semadeni} {et~al.}(1998){Vazquez-Semadeni}, {Canto}, \&
  {Lizano}}]{VazquezCantoLizano1998}
{Vazquez-Semadeni}, E., {Canto}, J., \& {Lizano}, S. 1998, \apj, 492, 596

\bibitem[{{V{\'a}zquez-Semadeni} {et~al.}(2010){V{\'a}zquez-Semadeni},
  {Col{\'{\i}}n}, {G{\'o}mez}, {Ballesteros-Paredes}, \&
  {Watson}}]{VazquezSemadeniEtAl2010}
{V{\'a}zquez-Semadeni}, E., {Col{\'{\i}}n}, P., {G{\'o}mez}, G.~C.,
  {Ballesteros-Paredes}, J., \& {Watson}, A.~W. 2010, \apj, 715, 1302

\bibitem[{{Veltchev} {et~al.}(2011){Veltchev}, {Klessen}, \&
  {Clark}}]{VeltchevKlessenClark2011}
{Veltchev}, T.~V., {Klessen}, R.~S., \& {Clark}, P.~C. 2011, \mnras, 411, 301

\bibitem[{{Vishniac}(1994)}]{Vishniac1994}
{Vishniac}, E.~T. 1994, \apj, 428, 186

\bibitem[{{Wang} {et~al.}(2015){Wang}, {Testi}, {Ginsburg}, {Walmsley},
  {Molinari}, \& {Schisano}}]{WangEtAl2015}
{Wang}, K., {Testi}, L., {Ginsburg}, A., {et~al.} 2015, \mnras, 450, 4043

\bibitem[{{Wang} {et~al.}(2010){Wang}, {Li}, {Abel}, \&
  {Nakamura}}]{WangEtAl2010}
{Wang}, P., {Li}, Z.-Y., {Abel}, T., \& {Nakamura}, F. 2010, \apj, 709, 27

\bibitem[{{Yusef-Zadeh} \& {Morris}(1987)}]{YusefZadehMorris1987}
{Yusef-Zadeh}, F., \& {Morris}, M. 1987, \apj, 322, 721

\bibitem[{{Zhang} {et~al.}(2014){Zhang}, {Qiu}, {Girart}, {(Baobab Liu},
  {Tang}, {Koch}, {Li}, {Keto}, {Ho}, {Rao}, {Lai}, {Ching}, {Frau}, {Chen},
  {Li}, {Padovani}, {Bontemps}, {Csengeri}, \& {Ju{\'a}rez}}]{ZhangEtAl2014}
{Zhang}, Q., {Qiu}, K., {Girart}, J.~M., {et~al.} 2014, \apj, 792, 116

\bibitem[{{Zhu} \& {Shen}(2013)}]{ZhuShen2013}
{Zhu}, Z., \& {Shen}, M. 2013, in IAU Symposium, Vol. 289, IAU Symposium, ed.
  R.~{de Grijs}, 444--447

\end{thebibliography}

\end{document}